%% file: pandey_mitra_perlekar.tex
\documentclass[]{jfm}
\usepackage{graphicx}
\usepackage{newtxmath}
\usepackage{relsize}
\usepackage{natbib}
\usepackage{hyperref}
\usepackage{textcomp}
\hypersetup{
    colorlinks = true,
    urlcolor   = blue,
    citecolor  = black,
}

\newcommand{\RomanNumeralCaps}[1]

\usepackage{bm}
\usepackage{color}
\usepackage{epstopdf}
\usepackage{float}
\usepackage{subfloat}
\usepackage[normalem]{ulem}
\definecolor{C0}{rgb}{0.12,0.47,0.705}
\definecolor{C1}{rgb}{1,0.498,0.055}
\definecolor{C2}{rgb}{0.173,0.55,0.173}
\definecolor{C3}{rgb}{0.839,0.153,0.157}
\definecolor{C4}{rgb}{0.580,0.404,0.741}
\input{symdef.inc}

\title{Turbulence modulation in buoyancy-driven bubbly flows}

\author{Vikash Pandey\aff{1}, Dhrubaditya Mitra\aff{2}
\and Prasad Perlekar \aff{1}} 
\affiliation{\aff{1} TIFR Centre for Interdisciplinary Sciences, Tata Institute of Fundamental Research,  Gopanpally, Hyderabad 500046, India
  \aff{2} Nordita, KTH Royal Institute of Technology and
Stockholm University, Roslagstullsbacken 23, 10691 Stockholm, Sweden}

\begin{document}
\maketitle

\begin{abstract}
    We present a Direct Numerical Simulation (DNS) study of buoyancy-driven
    bubbly flows in the presence of large scale driving that generates
    turbulence.  On increasing the turbulence intensity: (a) the bubble
    trajectories become more curved, and (b) the average rise velocity of the
    bubbles decreases.  We find that the energy spectrum of the flow shows a
    pseudo-turbulence scaling for length scales smaller than the bubble
    diameter and a Kolmogorov scaling for scales larger than the bubble
    diameter.  We conduct a scale-by-scale energy budget analysis to understand
    the scaling behaviour observed in the spectrum.  Although our bubbles are
    weakly buoyant, the statistical properties of our DNS are consistent with
    the experiments that investigate turbulence modulation by air bubbles in water.
\end{abstract}


\section{Introduction}
\label{sec:intro}

The flow of suspension of deformable objects (bubbles or droplets) is omnipresent in
a variety of natural and industrial processes
~\citep{mudde_rev_2005,eat10,riss18,said_rev_2019,mathai_rev_2020}.  The presence of
particles dramatically alters the rheological and thereby mixing properties of
flows~\citep{almeras_ijmf2019, almeras2015, rosti2018rheology,
rosti2018suspensions}.  A swarm of rising bubbles in an otherwise quiescent fluid,
at moderate volume-fraction, generates pseudo-turbulence studied by several
experiments and numerical simulations over the last three
decades~\citep{lance_1991,mudde_rev_2005,riss18, mathai_rev_2020,pandey_2020}. 

A more complex but ubiquitous scenario is where large scale external stirring that
generates turbulence is also present along with the bubbles
\citep{deckwer,joshi_08,mathai_rev_2020}. In the absence of bubbles, a nonlinear
transfer of energy (maintaining constant energy flux) from forcing to dissipation
range characterizes turbulence \citep{kol41, frisch,pope}. How does the presence of
bubbles modify this flow? The answer, in principle, depends on the ratio of the
bubble diameter to the dissipation scale, the bubble volume fraction, and its
density and viscosity contrast with the ambient fluid.

Experiments with large scale forcing that generates nearly homogeneous and isotropic
flows, at large Reynolds number, show that the presence of bubbles dramatically
alters the energy spectrum for scales smaller than the bubble diameter
\citep{vivek_2016,almeras2017}. Although the liquid velocity fluctuations have been
well-characterized, an understanding of the energy transfer mechanisms remain mostly
unexplored.

Direct Numerical Simulation (DNS) studies of bubbly flows have
explored: a) buoyancy-driven flows that generate pseudo-turbulence or
bubble induced agitation in the absence of external stirring
~\citep{bunner_tryg_2002,bunner_rvel_2002,roghair,pandey_2020,ramadugu_2020,chibbaro_2021}, 
b) modulation of turbulence by suspension of neutrally buoyant particles~\citep{ros_2019,you_2020},  
and c) Lagrangian investigations of an isolated bubble in the presence of external stirring~\citep{loisy_2017}.
However, to the best of our knowledge, a numerical study designed to unravel the statistical properties of 
buoyancy-driven bubbly flows in presence of external stirring is still missing.

Most numerical studies are restricted to low or moderate Galilei numbers because
extremely fine grids are required to fully resolve bubbles with high-density and
viscosity contrasts (for e.g., air bubbles in water)
\citep{cano-lozano_2016,chibbaro_2021}. Furthermore, the use of second-order
finite-difference methods limits the range of Reynolds numbers accessible to these
simulations~\citep{canuto}.

Fortunately, the DNS  studies of buoyancy-driven bubbly flow have shown that the 
statistical properties  of pseudo-turbulence such as the PDF of velocity
fluctuations, the scaling of the energy  spectrum, and the energy transfer
mechanisms are universal and do not depend upon  density and viscosity ratios
\citep{pandey_2020,ramadugu_2020,chibbaro_2021}. A key finding of these studies is
the presence of energy flux from length scales corresponding to the bubble diameter
to small scales.  This has also been confirmed in a recent study on bubble-laden
turbulent channel flow \citep{Ma_2021}.  Motivated by these findings, in this
article, we investigate turbulence modulation in suspensions of weakly buoyant
bubbles.  Similar to the experiments, we characterize the flow in terms of the
`bubblance' parameter $b=\Phi \paren{\Vzero/\uzero}^2$, where $\Phi$ is the bubble
volume fraction, $\Vzero$ is the rise velocity of an isolated bubble in a quiescent
fluid, and $\uzero$ is the r.m.s. velocity of the turbulent flow in the absence of
bubbles. The two extreme limits $b=0$ and $b=\infty$ correspond to pure fluid
turbulence and buoyancy-driven bubbly flow, respectively.

\section{Model}
We simulate the Navier-Stokes (NS) equations with a surface tension force to
investigate the suspension of bubbles.  Since we are interested in studying the
weakly buoyant regime, we invoke the Boussinesq approximation
\citep{chandra1981,pandey_2020} to get,
\begin{equation}
  D_t\uu = \nu \nabla^2\uu  -\nabla P +  \Fs +\Fg + \Ftrb,~{\rm and}~
  \nabla \cdot \uu = 0.
  \label{ns1}
\end{equation}
Here $\uu$ is the velocity field, $D_t \equiv \partial_t + \uu\cdot\grad$ is the
material derivative, $P$ is the
pressure field, and $\nu$ is the viscosity (assumed to be identical in the two
phases).  The two phases are distinguished using an indicator function $c$
which is equal to  $1$ in the liquid and $0$ inside bubble
\citep{popinet2018,Tryg2001}. The buoyancy force  $\Fg \equiv 2 \At [c - \ca]
{\bm g}$, where $\ca$ is the mean value of the indicator function,  $\At\equiv
(\rhof-\rhob)/(\rhof+\rhob)$ is the Atwood number,  ${\bm g}\equiv -g
\bm{\hat{z}}$ is the acceleration due to gravity, $\bm{\hat z}$ is a unit vector
along the vertical (positive $z$) direction, and $\rhof$ ($\rhob$) is the fluid
(bubble) density.  The surface tension force is  $\FF^\sigma \equiv \sigma
\kappa \nhat$, where $\kappa$ is the local  curvature of the bubble-front whose
unit normal is $\nhat$,  and $\sigma$ is the coefficient of the surface tension.
Turbulence is generated using a  large scale stirring force $\Ftrb$. For a
detailed discussion on the Boussinesq approximation, we refer the reader to
Appendix~\ref{app:bouss}. Experimentally small Atwood $\At$
(weakly buoyant regime) number flows can be realized in a mixture of oils
\citep{shu19,yi_toschi_sun_2021}.

We use a pseudo-spectral method \citep{canuto} for the DNS of \eqref{ns1} in a
periodic cube with each side of length $L\equiv 2 \pi$. The bubbles are resolved
using a front-tracking method. The same method had been earlier 
employed by us to investigate buoyancy-driven bubbly flows in absence of turbulent
stirring \citep{pandey_2020,ramadugu_2020}. For a detailed discussion on the
numerical implementation of front-tracking method to study a variety of multiphase
flows, we refer the reader to \citet{Tryg2001,popinet2018}.

For time-evolution, we use a second-order exponential time differencing
scheme~\citep{cox_2002} for \eqref{ns1} and a second-order Runge-Kutta scheme to
update the front. A substantial part of the computational effort is spent in
resolving the front; DNS with the bubbles is four times slower than the one without
them.  The large-scale stirring force is implemented in Fourier space, i.e.,
$\FFh^{\rm s} =
\ep
\uuh/\sum_{\kk} |\uuh|^2$ with $|\kk|\leq \kinj$
\citep{machiels_97,petersen_for_2010,perlekar_2019},  where $\uuh$ is the Fourier
transform of $\uu$ and $\kinj=2$.  This implementation ensures a constant rate of
energy injection, $\ep$.  

We discretize the simulation domain with $N^3$ collocation points, set
the initial velocity field such that the corresponding energy spectrum
$E(k,t=0)=\ep k^{4}\exp(-4k^2)$, and place $\Nb=80$ non-overlapping
spherical bubbles of diameter $d=0.46$ at random locations such that
no two bubbles overlap.

The dimensionless numbers that characterize the flow are the Taylor-scale Reynolds
number $\Rey \equiv u_0 \lambda/\nu$,   the Galilei number $\Ga\equiv\sqrt{2 \At g
d^{3}/\nu^2}$ the Bond number $\Bo \equiv {2\At \rho_a g d^2/\sigma}$,  and the
bubblance parameter $b\equiv \Phi \paren{\Vzero/\uzero}^2$, where $\Phi \equiv \Nb
(\pi/6) (d/L)^3$ is the volume fraction occupied by the bubbles, $\Vzero \approx
0.8$ is
the rise speed of a single bubble of diameter $d$ in quiescent fluid, \REM{%
$\delta\rho \equiv \rhof - \rhob$,} $\lambda\equiv \sqrt{15 \nu u_0^2/\ep}$ is
the Taylor-microscale, $\uzero \equiv \sqrt{2E/3}$ is the r.m.s. velocity in absence
of bubbles, $E \equiv \bra{\mid\uu\mid^2}/2$ is the average kinetic energy,  we set
the average  density $\rho_a=1$. The parameters used in our DNS are summarized in
table~\ref{tab:runs}. We conduct a grid-resolution study in
Appendix~\ref{app:sec2}  to show that our simulations are well resolved.

\begin{table}
    \begin{center}
        \begin{tabular}{lcccccccccc}
            runs & $N$  & $\Rey$ & $\epn\times10^{-2}$ & $\ep \times 10^{-2}$ &
            $\epg \times 10^{-3}$ &   $\lambda$ &  $\eta\times10^{-2}$ &  b \\
            ${\tt R0}$   & $720$ &  -- & $0.7$ &-- & $7.0$   & --  & --  & $\infty$ \\
            ${\tt R1}$  & $720$ &  $79$ & $0.9$ & $0.25$ & $6.8$ & $0.37$  & $2.0$  & 0.35 \\
            ${\tt R2}$ & $720$ &  $95$ & $1.2$ & $0.5$& $6.7$  & $0.34$ & $1.8$  &0.21  \\
            ${\tt R3}$  & $720$ &  $110$ & $1.5$ &$1.0$ &$6.0$ & $0.31$ & $1.5$ & 0.13 \\
        \end{tabular}
        \caption{\label{tab:runs} Parameters for our DNS runs ${\tt R0-3}$. Here,
            $\epn=\nu \bra{ |\nabla \uu|^2 } $ is the viscous dissipation rate,
            $\eta\equiv {(\nu^3/\ep)}^{1/4}$ is the Kolmogorov dissipation scale,
            $\lambda$ is the Taylor microscale, the energy injection rates due to
            large scale stirring  and buoyancy are  $\ep \equiv\bra{\uu\cdot\FF^{\rm
            s}}$
            and $\epg \equiv \braket{\uu\cdot\FF^{\rm g}}$, respectively.  The
            angular brackets denote spatio-temporal averaging in the statistically
            steady-state.  For all the runs $L=2\pi$, $d=0.46$, and the
            dimensionless numbers $\Ga= 302$, $\Bo =1.8$, $\At = 0.04$, and $\Phi=
            1.64\%$  are kept fixed.  We run simulations $\tt{R1-R3}$ at least for a
            period of $\approx 5\tauL$ in the
            steady-state, where $\tauL \equiv L/(2 \uzero)$ is the large eddy turn
            over time. The simulation $\tt{R0}$ runs for a period of $10 L/V_0$ in
            the steady-state. The
            values of $\Phi$, $\Ga$, $\Bo$, and $\Rey$ used in our study are
            comparable to those used in the
        experiments~\citep{vivek_2016,almeras2017}.} 
    \end{center}
\end{table}

\section{Results}

In what follows, we first investigate the statistical properties of bubbles  rising
in the turbulent flow, we then investigate the statistical properties of the fluid
velocity fluctuations.  Although we study turbulence modulation in the presence of
weakly buoyant bubbles, we show in the subsequent sections that the statistical
properties of the flow are in qualitative agreement with experiments that typically
have large density and viscosity contrast. Finally, we present the results for the
spectral properties of the flow by using a scale-by-scale energy budget analysis.

\subsection{Bubble trajectories and rise velocity}

For every bubble, we monitor the time evolution of its center-of-mass $\XX_{\rm
i}(t)$ after every $\delta t=0.08\tau_\eta$ time interval, where ${\rm i}$ denotes
the bubble index, and $\tau_\eta = \sqrt{\nu/\ep}$ is the Kolmogorov dissipation
time scale.  From the bubble tracks, we obtain the center-of-mass velocity $\VV_{\rm
i}(t)$ and the acceleration $\bm{A}_{\rm i}(t)$ using centered, second-order,
finite-differences.

\begin{figure}
  \centering
  \includegraphics[width=0.49\linewidth]{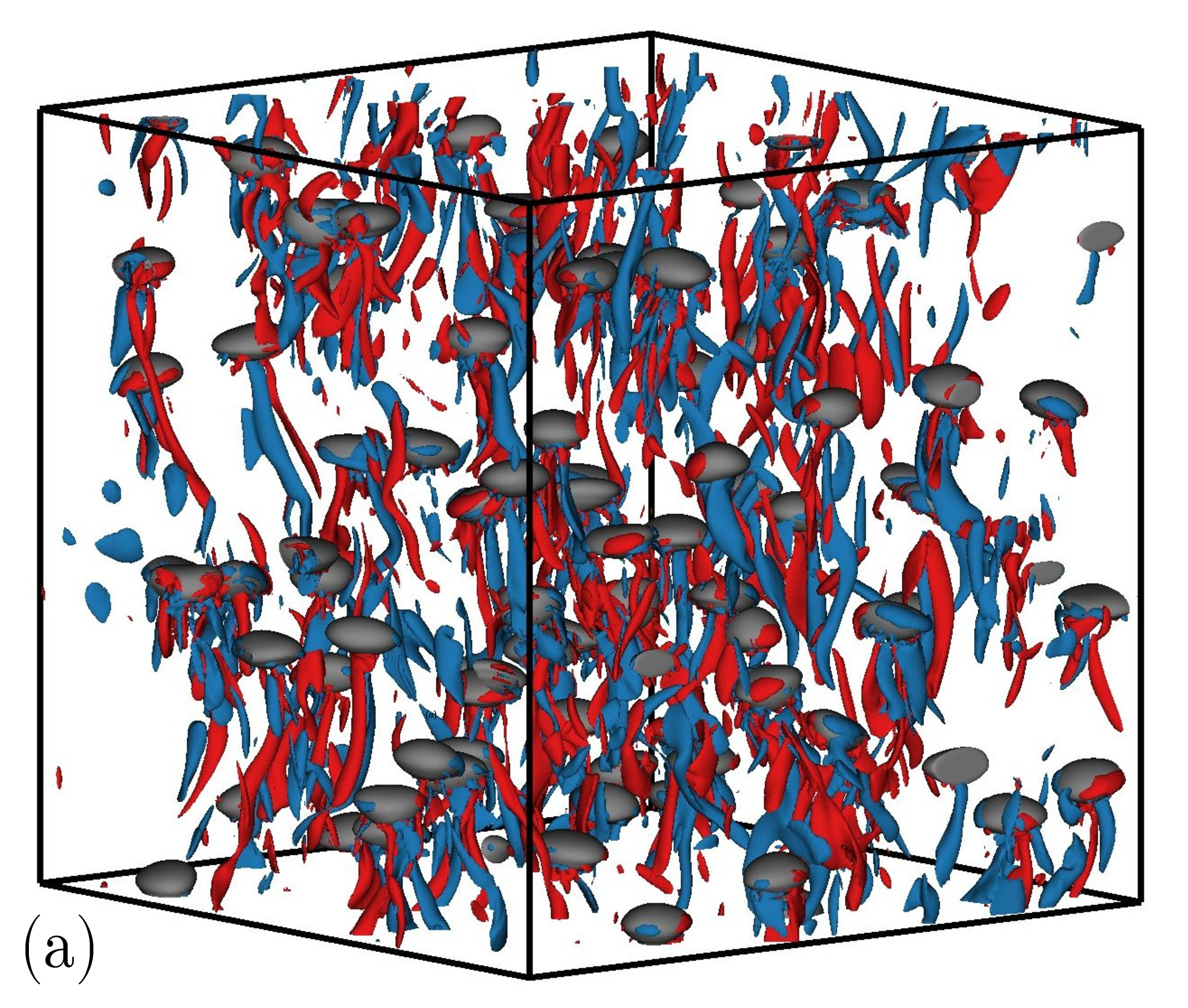} 
  \includegraphics[width=0.49\linewidth]{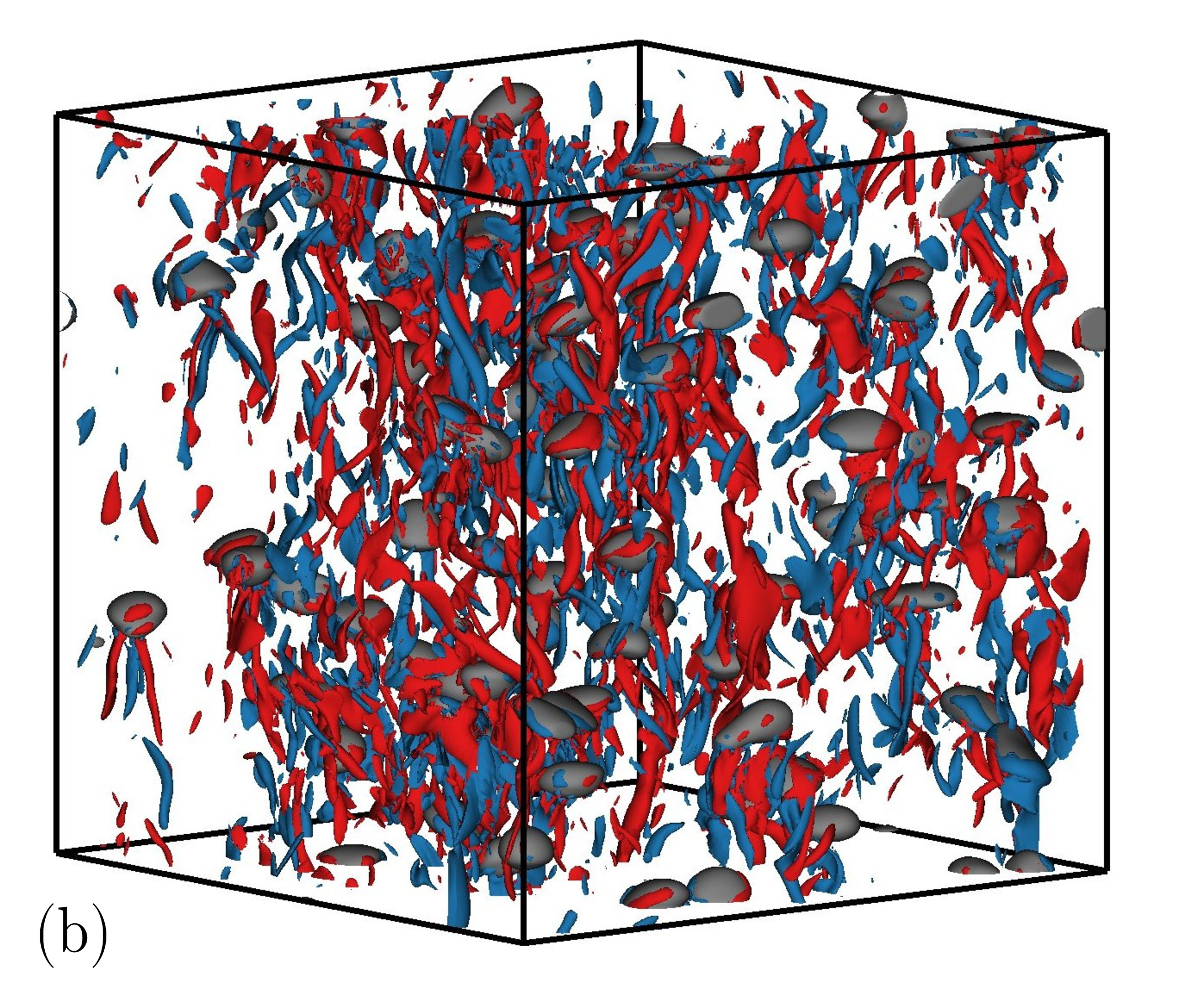} 
  \includegraphics[width=0.49\linewidth]{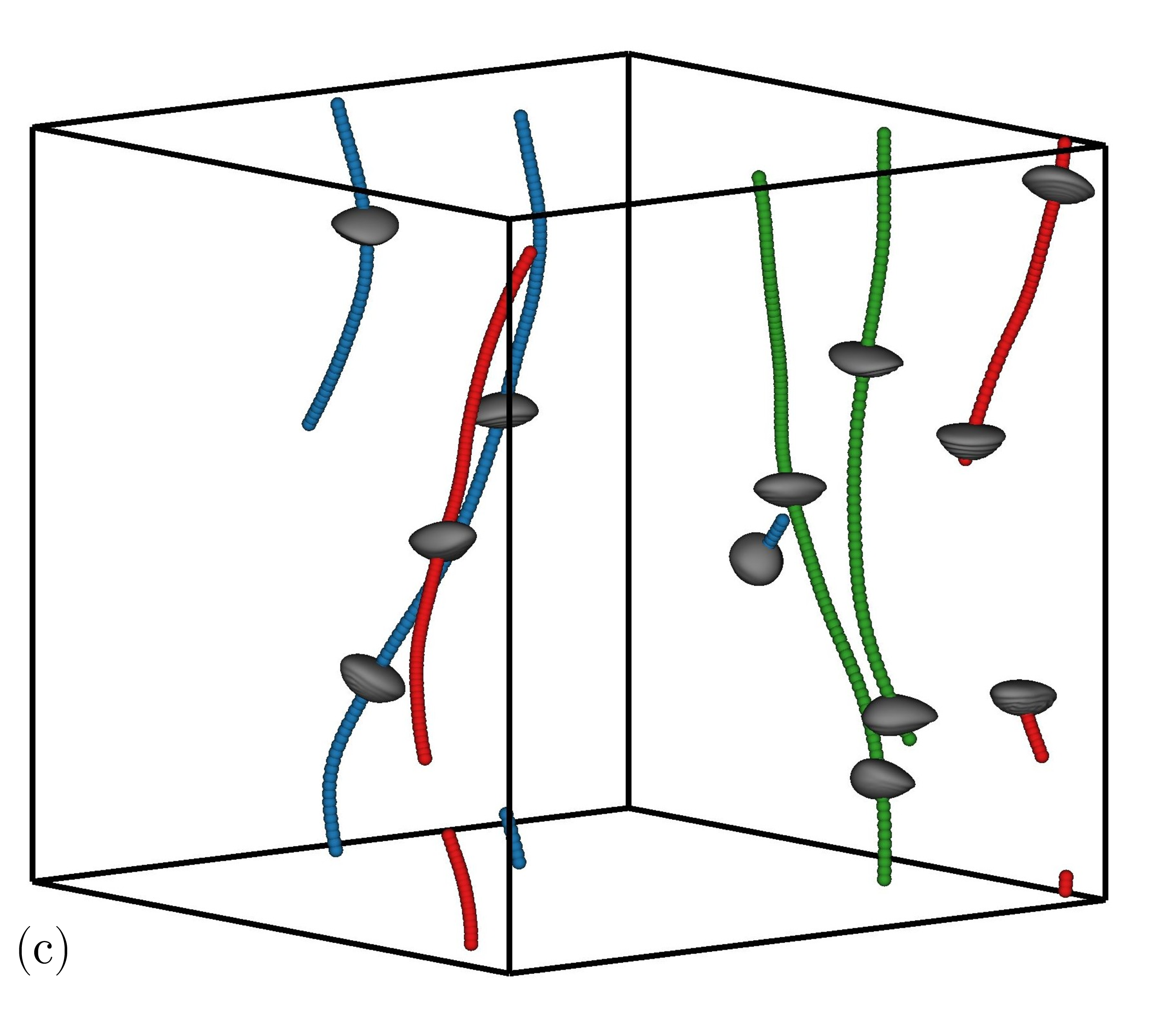}
  \includegraphics[width=0.49\linewidth]{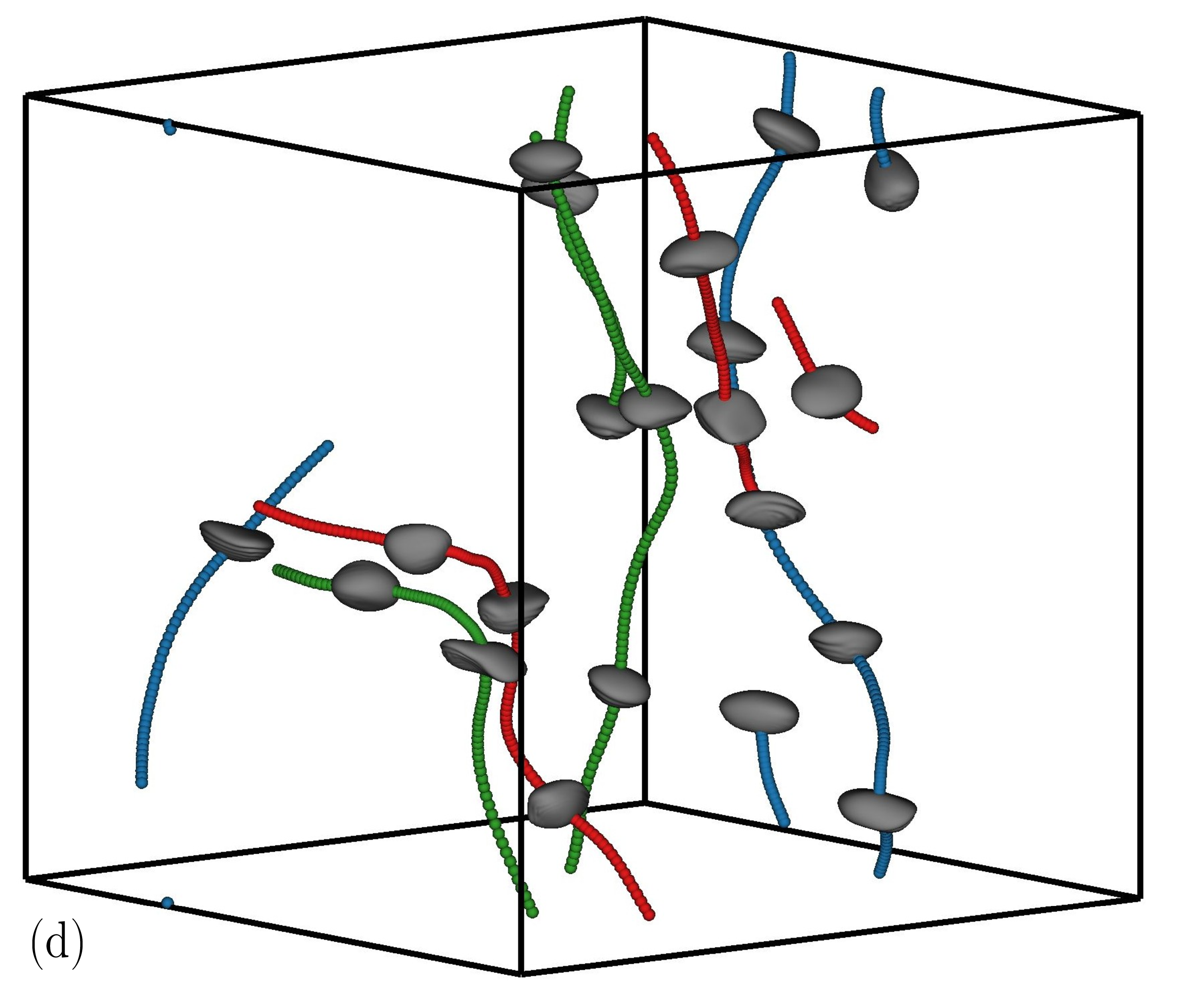}
  \includegraphics[width=0.49\linewidth]{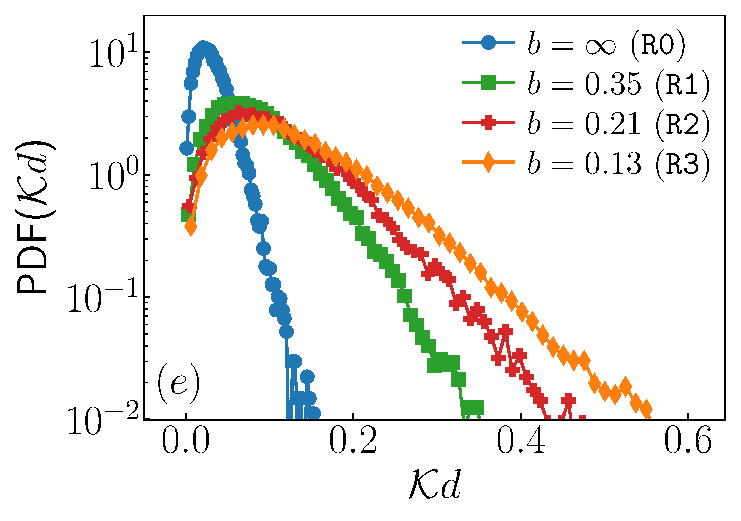}
  \includegraphics[width=0.49\linewidth]{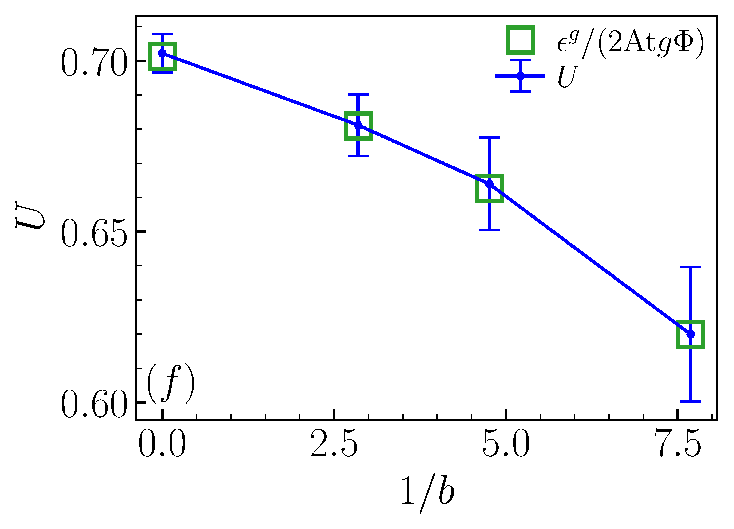}
  \caption{\label{fig:traj} Top panel: Representative steady-state
    snapshot of the bubbles and super-imposed iso-surfaces of the
    $z$-component of the vorticity field $\omega_z=\hat{\bm z} \cdot
    {\nabla \times \uu}$ for $\omega_z=\pm 3\bra{\omega_z^2}^{1/2}$ for
    (a) $b=0.35$, and (b) $b=0.13$. Middle panel: Typical trajectories
    of the center-of-mass of bubbles in a turbulent flow for (c) $\Rey=79,
    b=0.35~(\tt{R1})$ and (d) $\Rey=110,b=0.13~(\tt{R3})$.  Bottom panel:
    (e) The PDF of the curvature ${\mathcal K}$ for different values of $b$. (f) Plot
    showing that the bubble rise velocity increases with increasing $b$ or
    decreasing $\Rey$. We also show that $U$ obtained directly
    from the trajectories and the estimate $\epg/(2 \At g \Phi)$ are
    in excellent agreement.}
\end{figure}

The plots in \subfig{fig:traj}{a-b} show a representative snapshot of
bubbles and iso-vorticity surfaces for $\Rey=79,b=0.35$ and
$\Rey=110,b=0.13$, respectively.  In \subfig{fig:traj}{c-d} we show a
few typical trajectories for the same parameters. It is clear that
higher Reynolds number and small `bubblance' parameter corresponds to
more complex trajectories.  To quantify this behaviour we plot the
probability distribution function (PDF) of the curvature ${\mathcal
K}\equiv{\mid{\bm A} \times \bm{V}\mid/\mid\bm{V}\mid^3}$ in
\subfig{fig:traj}{e}. Consistent with the observation that the
trajectories are more curved for larger $\Rey$, we find that the
probability distribution function $P({\mathcal K})$ is broader-- has
an exponential tail.

Note that, \citet{bhatnagar2016deviation} showed that the PDF, of curvature of
trajectories of heavy inertial particles in homogeneous and isotropic turbulence, has a
power-law tail with an exponent of $-5/2$.  To the best of our knowledge no such results
exists for bubbles.

Another consequence of large-scale turbulent stirring is that the
average bubble rise velocity {$U \equiv (1/N_b) \sum_{{\rm i}=1}^{N_b}  \overline{\VV_{\rm i}(t)\cdot \bm{\hat z}}$ (see \subfig{fig:traj}{f})  increases with increasing $b$ (decreasing $\Rey$), where $\overline{(\cdot )}$ represents temporal averaging.}
 
In a recent study, \citet{sali_2020} show that the rise velocity of
the bubbles can be enhanced by turbulence provided the velocity ratio
$\gamma\equiv(\Vzero^2/(\ep d)^{2/3})<1$. Our DNS (see
Table~\ref{tab2}) and the experiments that investigate turbulence
modulation by bubbles \citep{lance_1991,vivek_2016} have $\gamma\gg1$.

Note that even for $b=\infty$, the rise velocity of a
bubble in a swarm is slightly smaller than the rise velocity of an
isolated bubble due to bubble-wake interactions \citep{risso_legendre_2010}.
Using the definition of $\Fg$ and noting that $\bra{u_z}=0$ in the Boussinesq regime, we
obtain $\epg= 2\At g \Phi U$ and verify it in \subfig{fig:traj}{f}. 

\begin{table}
    \begin{center}
        \begin{tabular}{lcccc}
            runs & ${\tt R1}$  & ${\tt R2}$ & ${\tt R3}$  \\
            $\gamma$   & $58.3$ & $36.7$  & $23.1$ \\
        \end{tabular}
        \caption{\label{tab2} Velocity ratio $\gamma$ for our DNS runs ${\tt R1-3}$} 
   \end{center}
\end{table}

  \subsection{Pair Distribution Function}
\begin{figure}
        \centering
        \includegraphics[width=0.31\linewidth]{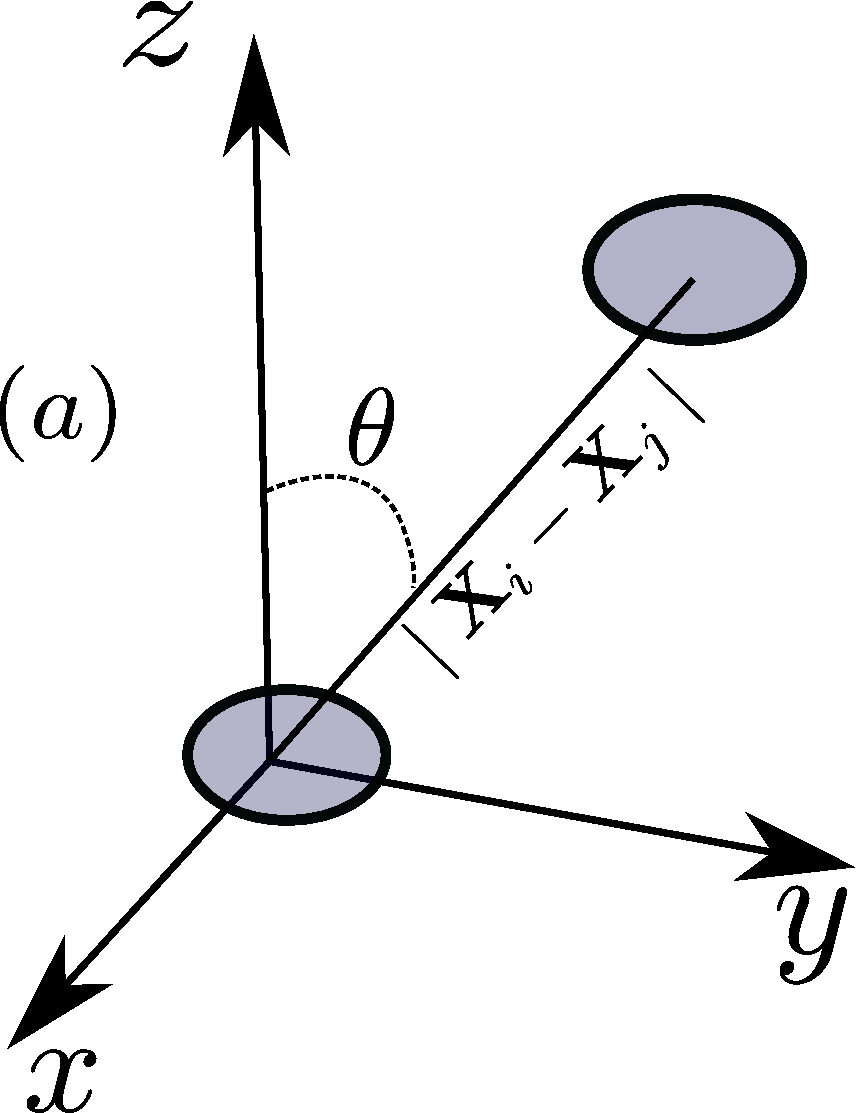}
        \includegraphics[width=0.48\linewidth]{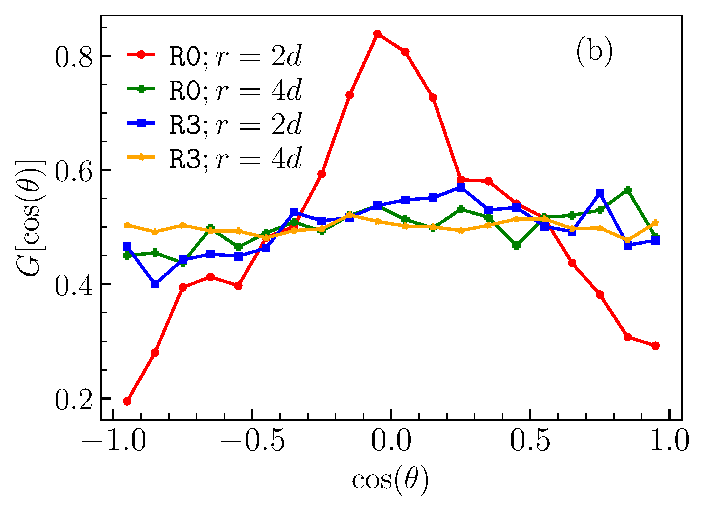}
	\caption{\label{fig:pair} {(a) The separation vector
${\bm r}= \bm{X}_{\rm i}-\bm{X}_{\rm j}$, and the angle $\theta$ between the
 $\bm{X}_{\rm ij}$ and $\hat{\bm z}$. The bubbles are represented as shaded
 ellipse. (b) The angular distribution function $G[r,\cos(\theta)]$ versus
 $\cos(\theta)$ for $r=2d$, and $r=4d$ in absence (presence) $b=\infty$ ($b=0.13$)
 of turbulence. The area under the curve is normalized to unity for each
 $G[r,\cos(\theta)]$ curve.}}
\end{figure}

    To understand the distribution of bubbles in the domain, following 
    \citet{bunner_rvel_2002}, we define the pair distribution
    function,

  \begin{align}
    G[r,\cos(\theta)] = {L^3\over
            \Nb(\Nb-1)}\overline{{\sum_{{\rm i}=1}^{\Nb}\sum_{{\rm j}=1, {\rm j} \ne {\rm i}}^{\Nb}
      \delta({\bm r} - \XX_{{\rm ij}},t)}},
    \label{eq:grtheta}
  \end{align}
where $\delta(\cdot)$ is the Dirac delta function, and $\XX_{{\rm ij}}=\XX_{\rm
i}-\XX_{\rm j}$.  In  \subfig{fig:pair}{a}, we sketch a bubble pair configuration to
show the co-ordinate system used for evaluating \eqref{eq:grtheta}. 
The plot of $G[r,\cos(\theta)]$ for $r=2d$ and $4d$ is shown in \subfig{fig:pair}{b}. At $b=\infty$, we observe a peak in $G[r,\cos(\theta)]$ for $r \approx 2d$ and $\cos(\theta)\approx 0$ indicating a
horizontal alignment of bubbles that are separated by a distance $2d$. Bubbles separated by distances, $r\geq 4d$ are uniformly distributed.  Our results are consistent with earlier 
 numerical studies of pseudo-turbulence \citep{bunner_rvel_2002,roghair_drag_2013}.  
In contrast, as turbulence makes flow more isotropic, for $b=0.13$ we find that $G[r,\cos(\theta)]$ is uniform which indicates that the bubbles  are uniformly distributed for all separations $r$.

\subsection{Average flow around a bubble}
  In this section, we study the average wake structure of the bubbles for
  different values of bubblance $b$.  \REM{In the statistically stationary state,
  we identify every bubble $i$ with its center-of-mass position $\XX_i$.}
  At a given time $t$, the velocity field in the center-of-mass frame of the 
  bubble ${\rm i}$ is given by 
\begin{align}
{\bm u}^{\rm CM}_{\rm i}({\bm \xi},t)= \uu({\bm \xi},t)- {\bm V}_{\rm i},
\end{align}
where ${\bm \xi}\equiv {\bm x}-\XX_{\rm i}$ and $-L/2 < (\xi_x,\xi_y,\xi_z) \leq L/2$.
The average flow around a bubble is then obtained by performing temporal averaging
over every  bubble as follows 
\begin{align}
\bm{u}^{\rm CM}({\bm \xi}) = \frac{1}{\Nb}\sum^{\Nb}_{{\rm i}=1} \overline{\bm{u}^{\rm CM}_{\rm i}({\bm \xi},t)}.
\end{align}

\begin{figure}
  \centering
  \includegraphics[width=\linewidth]{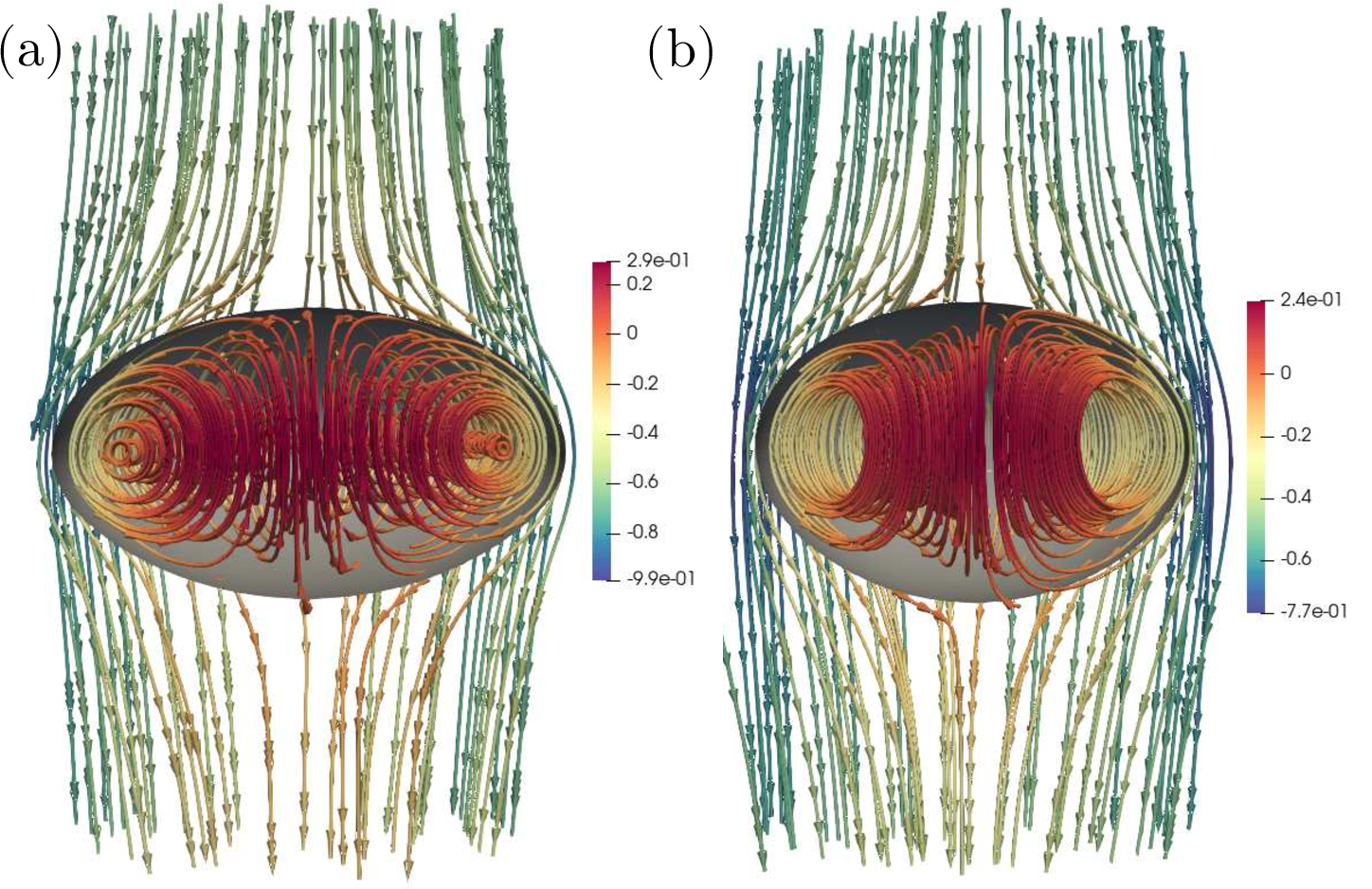}
  \caption{\label{fig:wake0} The streamline plot of the average
    velocity field in the frame of bubble for (a) $b=\infty$ run {\tt
R0} and (b) $b=0.13$ run
    {\tt R3}. The streamlines are colored according to $\uu^{\rm CM}\cdot\hat{\bm z}$}
\end{figure}

In \subfig{fig:wake0}{a,b} we plot the velocity streamlines of the
average velocity field $\bm{u}^{\rm CM}(\bm{\xi})$ for   $b=\infty$ ({\tt R0}) and 
$b=0.13$ ({\tt R3}).
Although the flow structure look qualitatively similar, we find that the 
bubble in the absence of large scale stirring is more ellipsoidal.
This can be understood by noting 
that presence of stirring
imposes stronger isotropy on the flow.

\begin{figure}
  \centering
  \includegraphics[width=0.45\linewidth]{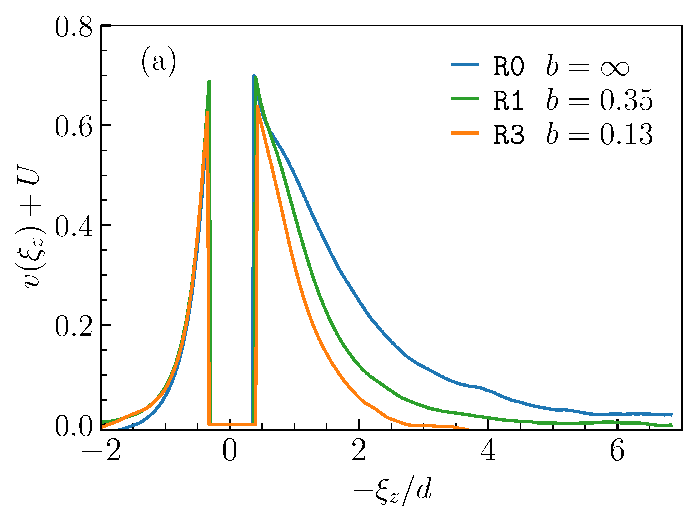}
  \includegraphics[width=0.45\linewidth]{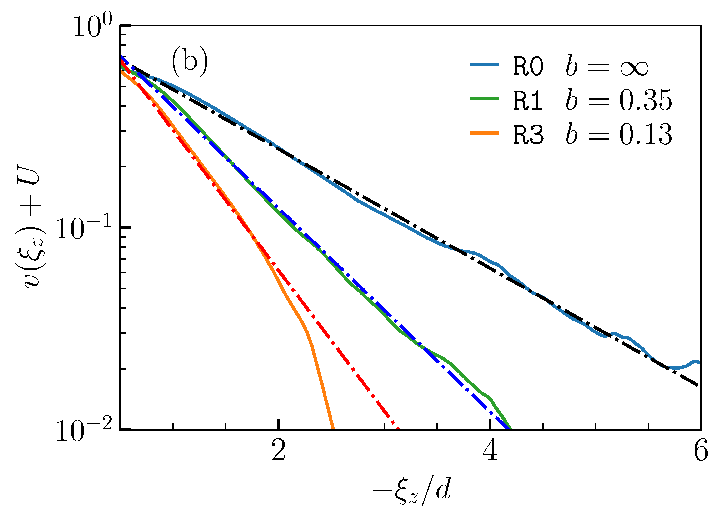}
  \caption{\label{fig:wake2} (a) The average bubble wake velocity
      $v(\xi_z)$ for run $\tt{R0}$ ($b=\infty$), $\tt{R1}$ ($b=0.35$), and $\tt{R3}$ ($b=0.13$). (b) Same as (a), but in semi-log scale to highlight the exponential decay of the velocity field in the wake region.  The dashed-dot line show the exponential fits $\sim\exp(-A z/d)$ to the data. We find $A=0.67,1.15$
and $1.6$ for $\tt{R0,R1}$, and $\tt{R3}$ respectively.}
\end{figure}

  To quantify, the behavior of the average bubble wake, similar to the
  experiments  \citep{risso_wake_2002,almeras2017}
  we plot $v(\xi_z)\equiv \bm{u}^{\rm CM}(0,0,\xi_z)\cdot\hat{\bm{z}}$ in
  \subfig{fig:wake2}{} and find
  that it decays exponentially $v(\xi_z)\sim C \exp(-A \xi_z/d)$ in the wake
  region for all values of $b$.
  However, consistent with earlier observations, presence of stirring leads to
  a faster decay of the wake.
  Therefore, for small $b$ (or large $\Rey$) we expect (see next section) the  velocity fluctuations to be similar to homogeneous, isotropic turbulence.

\subsection{Liquid velocity fluctuations}
\begin{figure}
  \begin{center}
    \includegraphics[width=0.45\linewidth]{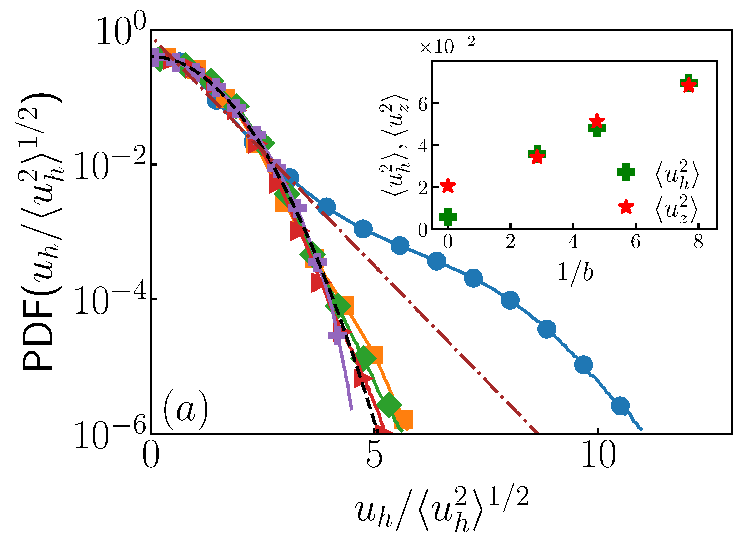}
    \includegraphics[width=0.45\linewidth]{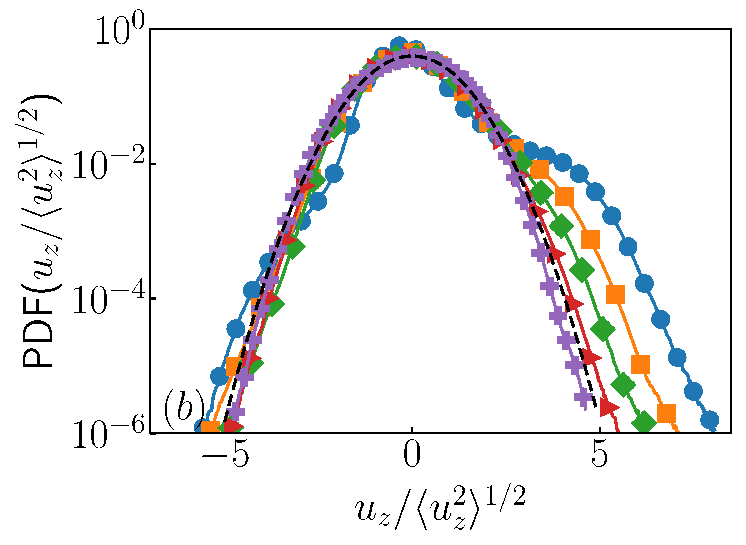}
    \caption{\label{fig:hist} The PDF of the {horizontal (a) and the
      vertical (b) component of the liquid velocity fluctuations} for different values of $b$
      [ $\textcolor{C0}{\large{\symb{\bullet}}}~b=\infty$ (${\tt R0}$),
      $\textcolor{C1}{\large{\symb{\blacksquare}}}~b=0.35$ (${\tt R1}$),
      $\textcolor{C2}{\large{\symb{\blacklozenge}}}~b=0.21$ (${\tt R2}$),
      $\textcolor{C3}{\large{\symb{\blacktriangleright}}}~b=0.13$ (${\tt
        R3}$), $\textcolor{C4}{\large{\symb{\boldsymbol{+}}}}~b=0$
      ($\Rey=110$)]. The black dashed line indicates a Gaussian
      distribution, and the brown dash-dot line in panel (a) shows the
      exponential distribution. (Inset) Variance of the horizontal and
      vertical velocity fluctuations increases with an increase in the
      stirring intensity $1/b$.}
  \end{center}
\end{figure}

The PDFs of the normalized horizontal and vertical liquid velocity
fluctuation with varying $b$ are shown in \Fig{fig:hist}. For
$b=\infty$, our results agree with the earlier studies on
pseudo-turbulence
\citep{risso_legendre_2010,risso_pdf_2016,pandey_2020}: the PDF of the
horizontal component shows exponential behaviour and the PDF of the
vertical component has a Gaussian core and is positively skewed.  The
presence of external stirring dramatically alters the PDFs as they
tend to a Gaussian distribution with decreasing $b$ (increasing
$\Rey$). Indeed, in {\subfig{fig:hist}{a,inset}} we verify that
$\langle u_h^2 \rangle \sim \langle u_z^2 \rangle$ on  {decreasing} $b$
confirming that the stirring makes the flow  isotropic. This is consistent with earlier 
experimental observations on turbulent bubbly flows \citep{vivek_2016,almeras2017}.

\subsection{Energy spectrum}
Earlier DNS studies \citep{roghair, pandey_2020,chibbaro_2021} have only
investigated the nature of the energy spectrum in the absence of large
scale turbulent forcing.  These studies, consistent with experiments,
confirm the presence of a $k^{-3}$ scaling in the spectrum that
appears because of the balance of net energy production in the wakes
with viscous dissipation.

Experiments have investigated temporal spectrum of the Eulerian liquid
velocity fluctuations in presence of a large scale stirring. They
observe a Kolmogorov spectrum for frequencies smaller than the bubble
frequency and a pseudo-turbulence scaling for higher frequencies
\citep{lance_1991,vivek_2016,almeras2017}.
 
Hence we expect that in our simulations we would find a Kolmogorov
scaling, for wavenumbers $k<\kd$, with a crossover to
pseudo-turbulence scaling for $k>\kd$, where $\kd\equiv 2\pi/d$ is the
wavenumber corresponding to the bubble diameter.

\begin{figure}
  \centering
      \includegraphics[width=0.45\linewidth]{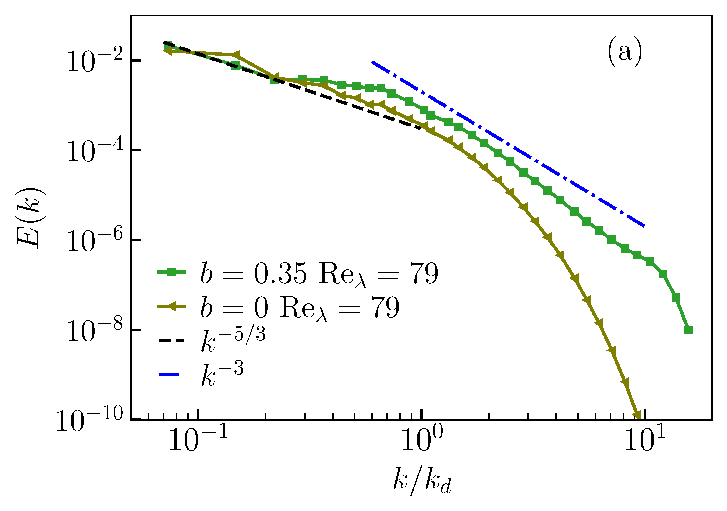}
      \includegraphics[width=0.45\linewidth]{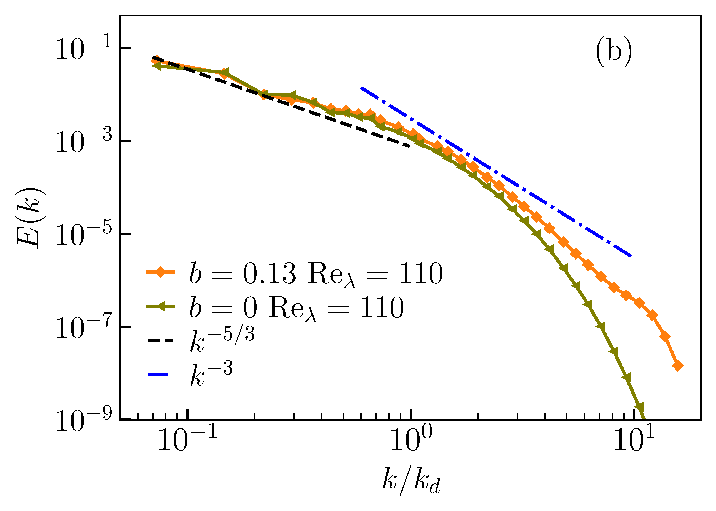}
     \caption{\label{fig:spec} Log-log plot of the kinetic energy spectrum,
       $E(k)$ versus $k/\kd$ for (a) $b=0.35,\Rey=79$ ({\tt R1}) and (b)
       $b=0.13,\Rey=110$ ({\tt R3}).}
 \end{figure}

\begin{figure}
  \begin{center}
    \includegraphics[width=0.45\linewidth]{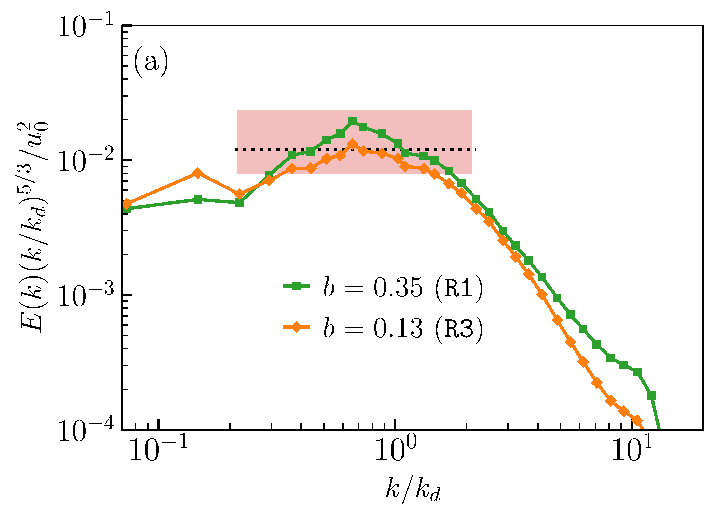} %
    \includegraphics[width=0.45\linewidth]{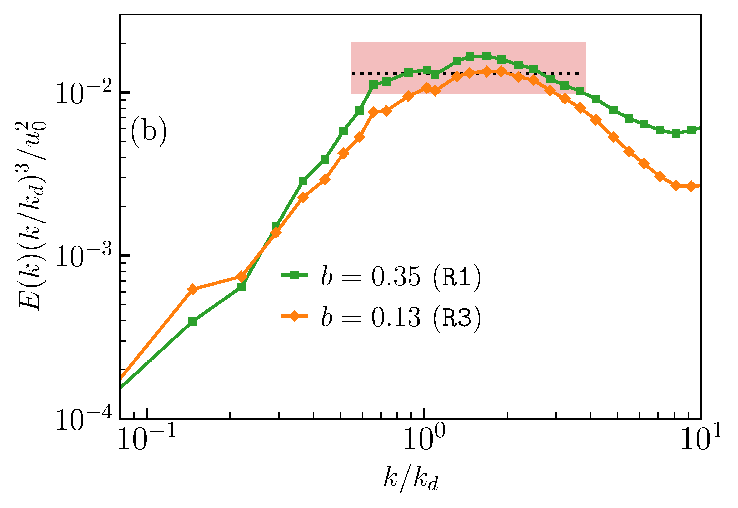} %
     \caption{\label{fig:spec_comp}  Compensated plot of the kinetic energy
spectrum highlighting the (a)$-5/3$, and (b) $-3$  scaling ranges. Horizontal dashed line and 
the shaded region indicate the scaling range.}
   \end{center}
 \end{figure}

In \Fig{fig:spec}, we plot the scaled energy spectrum for different
values of $b$ ($\Rey$).  As expected, we observe Kolmogorov scaling
$E(k)\sim k^{-5/3}$ for $k<\kd$ and a pseudo-turbulence scaling $E(k)
\sim k^{-3}$ for $k>\kd$. {In \subfig{fig:spec_comp}{a,b} we
plot the compensated  spectrum to highlight the region showing $-5/3$ and $-3$ scaling.} Note
that none of the scaling ranges are large enough to make an accurate
determination of the scaling exponent possible.  \REM{As $b$ decreases, the
range of $k$ over which the the pseudo-turbulence scaling holds
decreases, eventually for $b=0$ the energy spectrum becomes that of
homogeneous and isotropic turbulence.}

\subsection{Scale-by-scale energy budget and flux}
To lay bare the mechanism by which bubbly turbulence emerge we study
the scale-by-scale energy budget. Following \citet{pope} we define a
low-pass filtered velocity field coarse-grained at scale $\ell=2\pi/K$
as

  \begin{align}
    \uK(\xx) \equiv \int\exp(i\qq\cdot\xx) \GK(\qq)\uuh(\qq) {\rm d}\qq \/, {\rm with~} 
    \GK(\qq) \equiv \exp\left(-{\pi^2q^2\over 24 K^2}\right) \/.
          \label{eq:filter}
  \end{align}

Note that \citet{frisch,pandey_2020} use a sharp stepdown function as
a filter: $\GK(\qq) = 1$ for $\mid\qq\mid \leq K$ and zero otherwise,
whereas we use a smooth Gaussian filter \citep{pope}.  In what
follows, we use the symbol $\paren{\cdot}^{<}_{ K}$ to denote the
filtering operation \citep{frisch}.  In real space, this corresponds to
  \begin{align}
  \nonumber
    \uK(\xx) = \int G_\ell(\bm{r})\uu(\xx - \bm{r}){\rm d}\bm{r}\/,{\rm with~}
    G_\ell(\bm{r}) = \paren{{6 \over \pi\ell^2}}^{1\over 2}\exp\paren{-{6r^2\over \ell^2}}{~\rm{and}~}\ell \equiv 2\pi/K.\/
  \end{align}
Using the filtered velocity field, we obtain the following scale-by-scale energy budget
equation from \eqref{ns1}
\begin{align}
  \PiK + \FsK =   - \DK   + \FgK + \FturbK \/.
  \label{ebud}
\end{align}
Here $\FsK \equiv \bra{ \uK\cdot\fil{\FF^{\sigma}} }$ is the contribution from
surface-tension forces, $\FgK \equiv \bra{ \uK\cdot\fil{\FF^{\rm g}} }$ is the
contribution from buoyancy, and $\FturbK \equiv \bra{ \uK\cdot\fil{\Ftrb} }$ is the
contribution due to large-scale forcing.  To obtain the contribution from the nonlinear
term and viscous dissipation, following
\citet{eyink_1995,orszag_1998,pope}, we define a filtered version of
the Reynolds stress tensor,
\begin{align}
  \mT^{\ab}_{ K}({\bm x}) \equiv \fil{u^{\alpha}u^{\beta}} - \fil{u^{\alpha}}\fil{u^{\beta}}\/,
  \label{eq:reyst}
\end{align}
the rate-of-strain tensor
\begin{align}
  \mSK^{\ab}({\bm x}) \equiv \frac{1}{2}\left[\fil{\dela\ub}+\fil{\delb\ua} \right] \/, 
 \label{eq:rst}
\end{align}
and the local nonlinear energy flux
\begin{align}
  \piK({\bm x}) \equiv  -\mT^{\ab}_{ K}\mSK^{\ab} \/. 
  \label{eq:piK}
\end{align}
Using \eqref{eq:filter},\eqref{eq:reyst},\eqref{eq:rst}, and \eqref{eq:piK}, we get the 
net nonlinear flux $\PiK \equiv \bra{\piK}$, and %
the viscous contribution to the budget 
$\DK \equiv  2\nu \bra{\mSK^{\ab}\mSK^{\ab}}$ which is always positive.

\begin{figure}
    \begin{center}
        \includegraphics[width=0.45\linewidth]{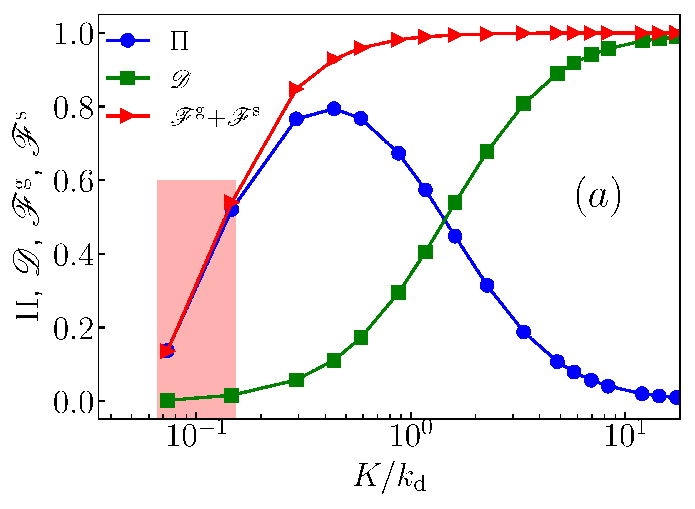}
        \includegraphics[width=0.45\linewidth]{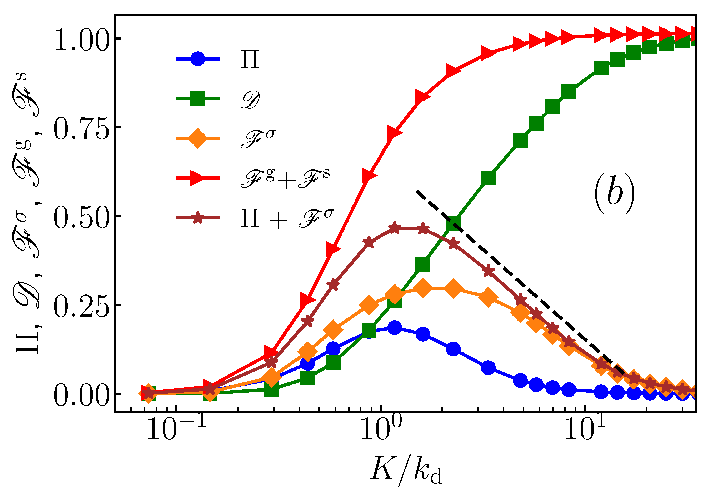}
        \includegraphics[width=0.45\linewidth]{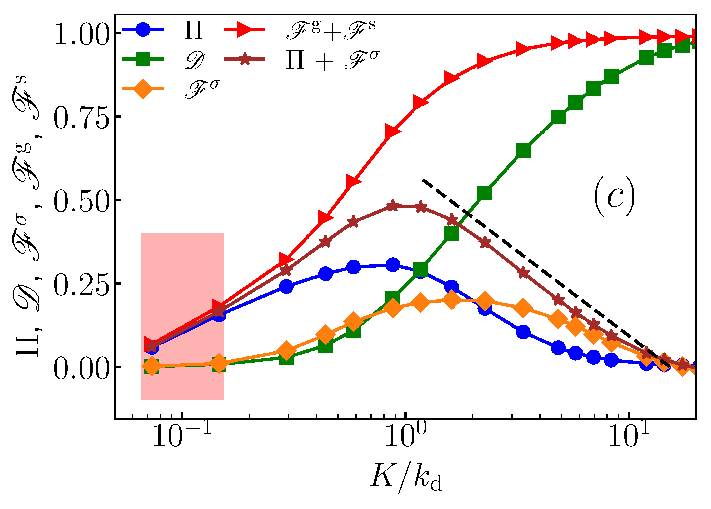}
        \includegraphics[width=0.45\linewidth]{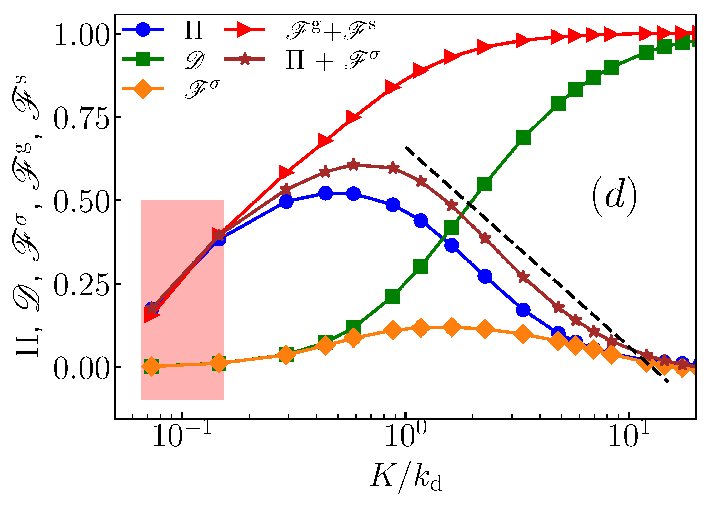}
        \caption{\label{fig:flux} Scale-by-scale energy budget: plot of the
        energy flux $\PiK$, cumulative viscous dissipation $\DK$,
        the surface tension contribution $\FsK$, the cumulative
        energy injected due to buoyancy $\FgK$, and the energy
        injected due to turbulent forcing $\FturbK$ for $b=0~(\Rey = 110)$
        (a), $b=\infty$ (b), $b=0.35$ (c), and $b=0.13$ (d). The black dashed
        line indicates $\log({K})$ scaling. In (a-d) we normalize the ordinate by the 
        viscous dissipation $\epn$. {In panel (a), (c) and (d) we
          mark the injection wavenumbers by a shaded region.}}
    \end{center}
\end{figure}


\begin{figure}
  \begin{center}
    \includegraphics[width=0.6\linewidth]{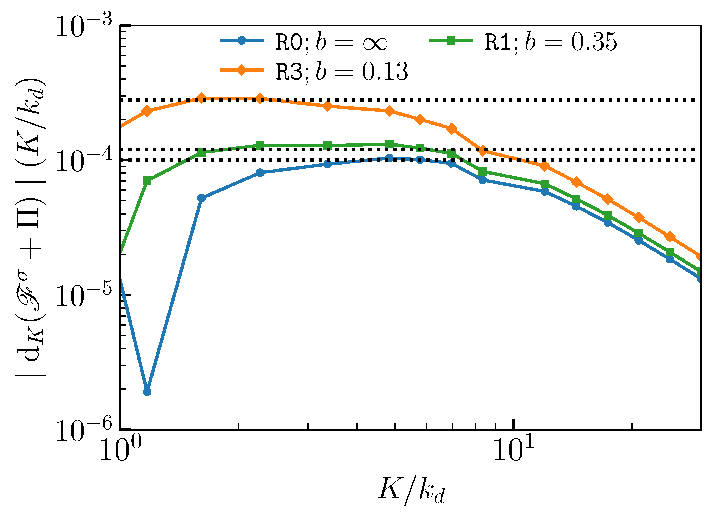}
    \caption{\label{fig:derflx} Log-log plot of $(K/k_d) |d(\FsK + \PiK)/dK|$ versus $K/k_d$
     for different values of the bubblance parameter $b$. Horizontal dashed lines  represent  $K^{-1}$ scaling. }
  \end{center}
\end{figure}

\subsubsection{Scale-by-scale energy budget in the absence of bubbles: $b=0$}
In this case buoyancy makes no contribution to the fluxes and \eqref{ebud} simplifies to 
\begin{equation}
  \PiK  =   - \DK + \FturbK \/.
  \label{eq:fluxbzero}
\end{equation}
The plot in \subfig{fig:flux}{a} shows the energy budget for $b=0$ ($\Rey=110$). 
Since the stirring force is limited to small Fourier modes 
$k\leq \kinj$, $\FturbK=\ep$ is a constant for $K>\kinj$.
The viscous contribution $\DK$ is significant only for very large $K \geq \keta$.
Hence, for intermediate values of $K$ in the inertial range ($\kinj < K < \keta$), the flux 
$\PiK = \FturbK$ remains a constant. 
The  four-fifth law of Kolmogorov and the Kolmogorov scaling, $E(k) \sim k^{-5/3}$, 
is a consequence of this constancy of flux~\citep[see, e.g., ][section 6.2]{frisch}.
Because of the moderate $\Rey=110$ used by us, the range of wavenumbers over which the 
flux is constant is very small.  A significant range of constant flux
is observed in very high $\Rey$ and large resolution
DNS~\citep{ishihara2009study}.

\subsubsection{Scale-by-scale budget in the absence of stirring ($b=\infty$):}
Next, in \subfig{fig:flux}{b} we study the other extreme, $b=\infty$.
Stirring makes no contribution here. Energy injection by buoyancy
forces happens around the scale of the bubble diameters, the flux due
to buoyancy $\FgK$ becomes almost a constant for $K \gg \kd$.  Hence
for $K \gg \kd$ we obtain
\begin{equation}
  \PiK + \FsK =   - \DK   + \FgK  \/,
  \label{eq:flux_binfty}
\end{equation}
with $\FgK$ approximately a constant. By taking a derivative of both sides of \eq{eq:flux_binfty}
with respect to $K$ at $K = k$ we obtain
\begin{equation}
	\left.\frac{d(\PiK+\FsK)}{dK}\right\rvert_{K=k} = \nu k^2E(k).
\label{eq:dflux}
\end{equation}
Our DNS shows that the net production 
$\PiK+\FsK \sim \log(K)$ \citep{lance_1991,pandey_2020}.
{Although, taking derivative can enhance approximation errors, we directly confirm the scaling relation in \Fig{fig:derflx}.
Generalizing \citet{lance_1991} argument if we now assume locality of net transfer then by dimensional
analysis  $\d(\Pi_K + \FsK)/\d K\mid_{K=k} \sim k^{-1}$ follows.}
Substituting in \eqref{eq:dflux} we obtain $E(k) \sim k^{-3}$ -- 
the spectrum of pseudo-turbulence \citep{lance_1991,martinez_2010,vivek_2016,almeras2017,bunner_tryg_2002,roghair,pandey_2020,ramadugu_2020}.
\citet{rissou_2011} has shown that the same $k^{-3}$-spectrum can be obtained,
under certain conditions, as a sum of localized random, statistically independent,
bursts; which comes from localized velocity disturbances caused by the bubbles.

\subsubsection{Scale-by-scale budget in the presence of both bubbles and stirring}
In \subfig{fig:flux}{c,d} we plot the energy budget for the two intermediate cases with 
$b=0.35$ and $b=0.13$.  
For $K \ll \kd$ both the buoyancy force and the surface tension contribute very little
to the flux. The viscous contribution is also very small as $\kd < \keta$, the dissipation wavenumber.
Let us also assume that there is a scale separation between the stirring scale, $\kinj$ and $\kd$,
with $\kinj \ll \kd$. 
Then for range of scales $\kinj < K < \kd $ the flux balance
gives $\PiK = \FturbK$, equal to a constant. 
Consequently we obtain $E(k) \sim k^{-5/3}$ for $\kinj <k < \kd$.
Next we consider $K \gg \kd$: the net contribution from both stirring and buoyancy forces 
$\FturbK + \FgK$ 
is almost a constant, hence we again obtain \eq{eq:dflux}. 
{Our DNS show that for both the bubblance, $b=0.35$, and
  $0.13$, $\PiK+\FsK \sim \log(K)$ (see \Fig{fig:derflx}). Although their individual
  contribution to the energy budget does depend on $b$,
in particular: for $b=0.13$, $\PiK$ is larger than $\FsK$, but for $b=0.35$, $\PiK$ is smaller than $\FsK$.}
\REM{Our DNS shows that for both of these cases
$\PiK+\FsK \sim \log(K)$ although their individual contribution does depend on $b$. 
In particular: for $b=0.13$, $\PiK$ is larger than $\FsK$, but for $b=0.35$, $\PiK$ is smaller than $\FsK$.} 
Hence for both of these cases we obtain $E(k) \sim k^{-3}$ for $ k >
\kd $ and $E(k)\sim k^{-5/3}$ for $k<\kd$.

In Appendix \ref{app}, we show that qualitatively similar results are obtained even by using a 
sharp filter instead of a Gaussian filter.
\subsubsection{Spatial distribution of the nonlinear energy flux $\piK(x)$}
For homogeneous and isotropic turbulence, for any $K$ in the inertial
range, the net nonlinear 
flux $\PiK$ is positive, i.e., on average energy flows from small to large $K$ or from large 
to small spatial scales. 
Kraichnan \citep{krai_1974,eyink_1995} argued that the local nonlinear energy flux $\piK$ \eqref{eq:piK} satisfies the refined similarity hypothesis. 
Using DNS, \citet{chen_2003} verified this and showed that the scaling exponents of the flux 
show multiscaling. 
The multiscale analysis of the flux is also crucial to model 
subgrid scale dissipation in large-eddy simulations \citep{meneveau_katz_2000}.

To the best of our knowledge, the spatial distribution of local energy 
flux in bubbly flows remains unexplored. 
How does the sign of this flux correlate with the bubbles?  
For example, is the flux pre-dominantly positive in the wake of a bubble?  
In the following discussion, we address this question by performing a multiscale 
analysis of the local nonlinear energy flux $\piK(\xx)$ with varying filtering 
scale $\ell \sim 1/K$.

\begin{figure}
    \centering
    \includegraphics[width=1\linewidth]{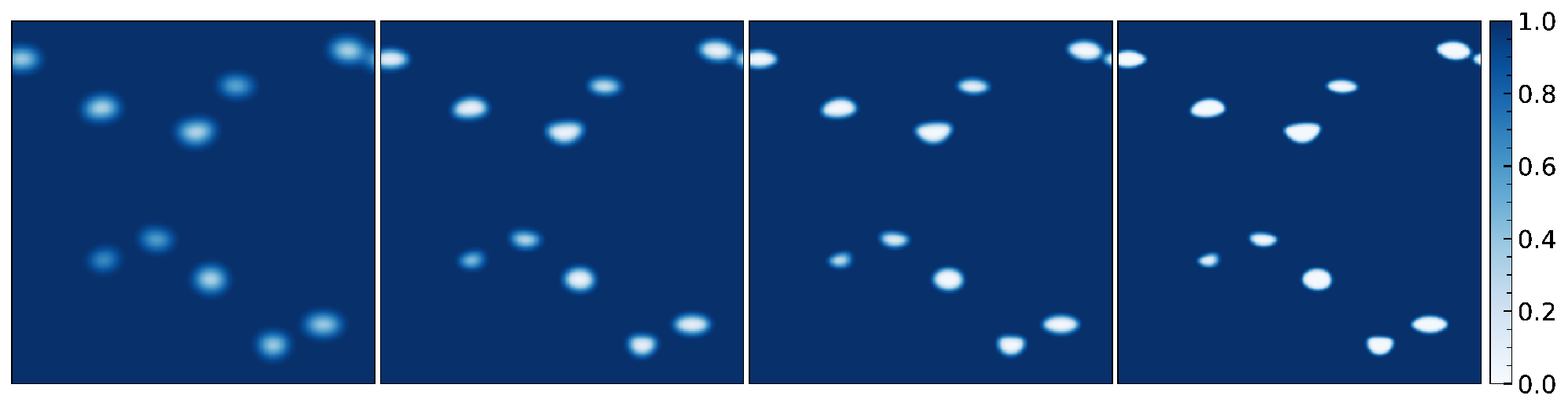}
    \includegraphics[width=1\linewidth]{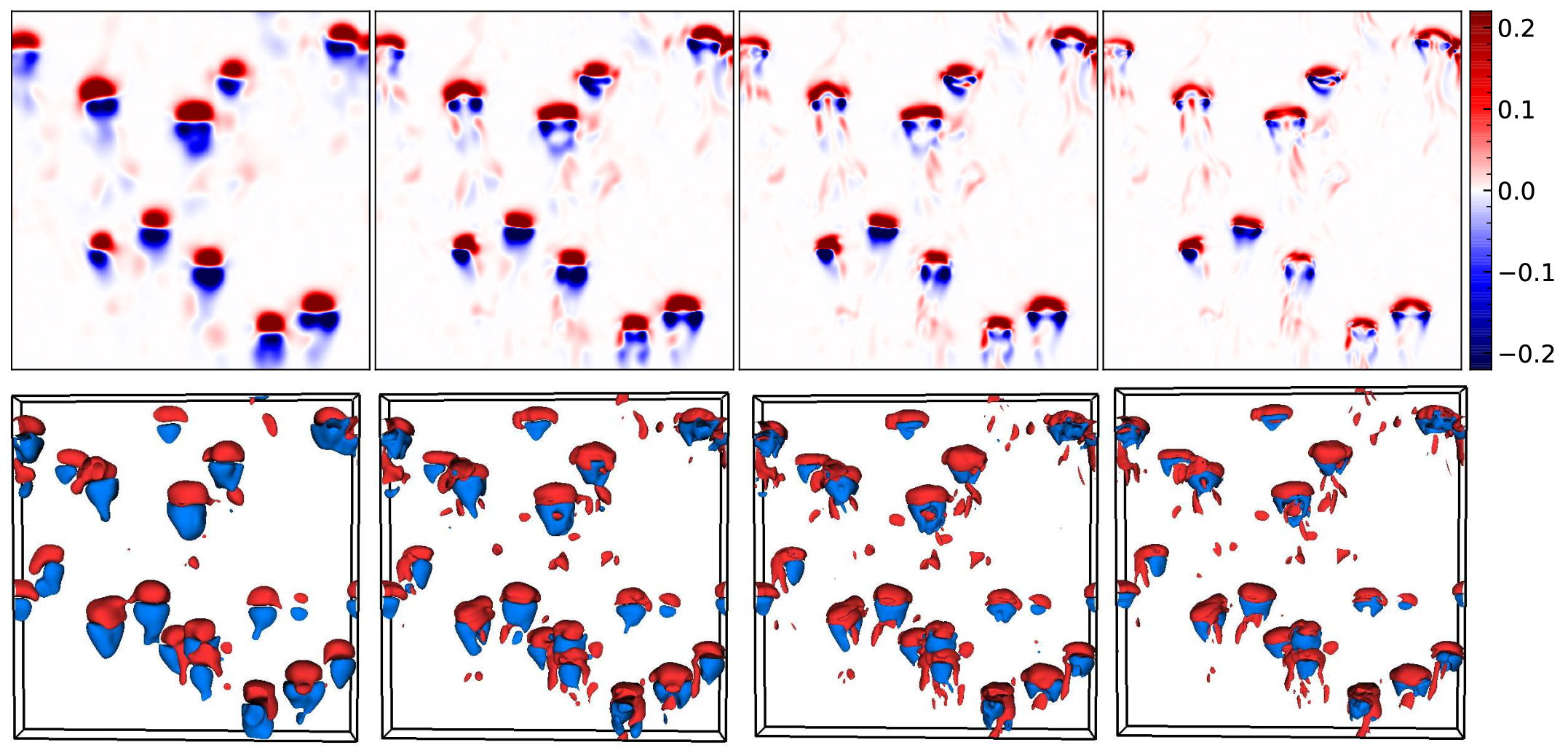}
    \caption{\label{flx_snap_R0} Buoyancy driven flow in absence of
      stirring ($b=\infty,{\tt R0}$).  The pseudocolor plot of the filtered
      indicator function $c$ (top panel) and the local nonlinear flux
      $\piK/\max(\pi_{\kd})$ (middle panel) in the $y=L/2$ plane. Constant-$\piK$
      isosurfaces for $|\piK|=0.03 \max(\pi_{\kd})$ in a slab $L \times 2d \times L$ 
      around the $y=L/2$ plane (bottom
      panel). The filter wavenumber (scale) is increased (decreased) from
      left to right $K/\kd=0.6, 1.0, 1.4$ and $2.2$.}
\end{figure}

In \Fig{flx_snap_R0}, we show a typical snapshot from the run with no external stirring, $b=\infty$. 
The position of the bubbles is shown by plotting the indicator function in the top panel. 
In the middle and bottom panel, we plot the local nonlinear flux $\piK$. 
In each panel, we use four different 
values for the filtering wavenumber $K/\kd = 0.6, 1.0, 1.4$, and $2.2$, from left to right. 
Note that we use a Gaussian filter; therefore, a proper distinction between liquid and 
bubble phase can be made only for $K>\kd$. 
We make the following observations:
\begin{enumerate}
\item In the front of the bubble, the energy is primarily transferred downscale, 
i.e., to scales smaller than $\ell \sim1/K$. 
\item Depending on the filtering scale, we observe both upscale and downscale transfer 
of energy in the wake of the bubble. 
For large $K$ (small $\ell$), downscale transfer of energy dominates the wake region, 
but there are also regions of upscale transfer.
\item On reducing the filter wavenumber $K$ (large $\ell$), we observe that the region of 
upscale transfer are enhanced in the aft region of the bubble. 
For the smallest filtering wavenumber $K/\kd=0.6$, the front-aft region of the bubble 
has a similar structure but appears with opposite signs.
\end{enumerate}

\begin{figure}
    \centering
    \includegraphics[width=1\linewidth]{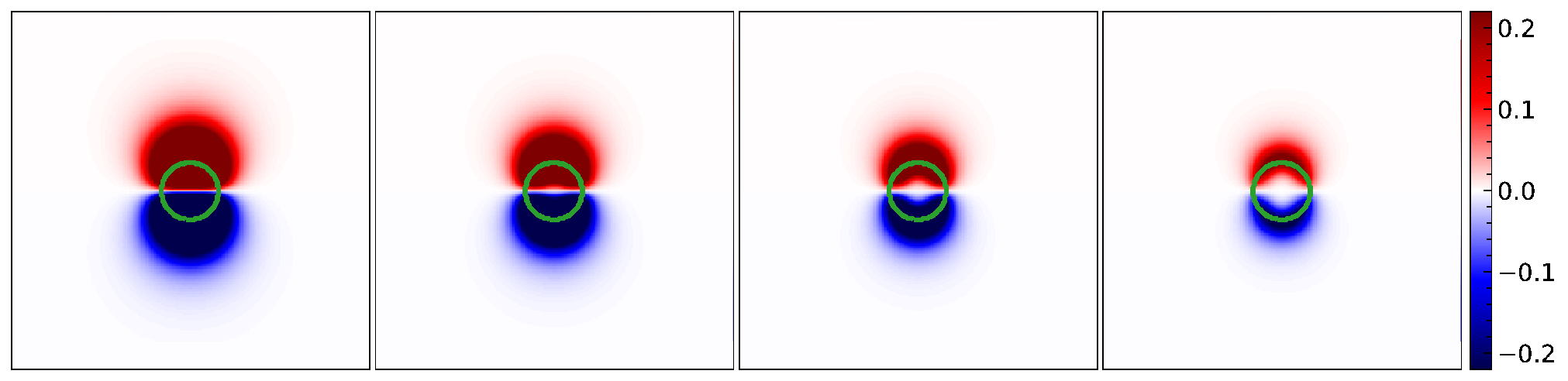}
    \caption{\label{flx_had} The space-dependent nonlinear flux $\piK/\max(\pi_{\kd})$ in the $y=L/2$ plane for the 
Hadamard-Rybczynski flow \Eq{eq:hadamard}.  The filter wavenumber (scale) is increased (decreased) from 
left to right  $K/\kd=0.6, 1.0, 1.4$ and $2.2$. The green line
  represents the bubble interface. }
\end{figure}

We can understand the fore-aft structure of the energy flux in the vicinity 
of a bubble in a straightforward manner.  
Consider a Stokesian spherical bubble with the same viscosity  as
  ambient fluid rising in a quiescent flow;  
the stream function is given by the 
Hadamard-Rybczynski solution \citep{hadamard1911,rybczynski1911,clift1978}:
\begin{equation}
\label{eq:hadamard}
\Psi(r,\theta) = {\Vzero r^2 \sin^2(\theta) \over 2}
\begin{cases}
-\left(1 - \dfrac{5 d}{8r} +
        \dfrac{d^3}{32 r^3} \right),~&\text{for}\ r\geq d/2,\\
\dfrac{1}{4}\left(1 - \dfrac{4 r^2}{d^2} \right)~&\text{for}\ r<d/2.
\end{cases}
\end{equation}
The radial and the angular component of the velocity field are  
$u_r=\partial_\theta \Psi/r^2 \sin(\theta) $ and 
$u_\theta=-\partial_r \Psi/r \sin(\theta)$. 
Using  \eq{eq:hadamard}, we calculate the nonlinear flux $\piK$ and plot it in  \Fig{flx_had}  
for four different values of the filtering wavenumber $K/\kd = 0.6, 1.0, 1.4$, and $2.2$. 
There is {a} downscale energy transfer in the front
and a upscale energy transfer at the back side 
of the bubble. Note that the net energy flux $\PiK$ is zero for the 
Hadamard-Rybczynski solution.
Comparing \Fig{flx_snap_R0} with \Fig{flx_had}, it seems that the
spatial distribution of the energy flux comprises of a
Hadamard-Rybczynski-like solution superimposed with turbulent
fluctuations generated in the wake region of a rising bubble.  Thus
our multiscale analysis of the spatial energy flux provides a direct
evidence that
the net forward energy flux in {\subfig{fig:flux}{b}}  is due to the bubble wakes.

\begin{figure}
    \centering
    \includegraphics[width=1\linewidth]{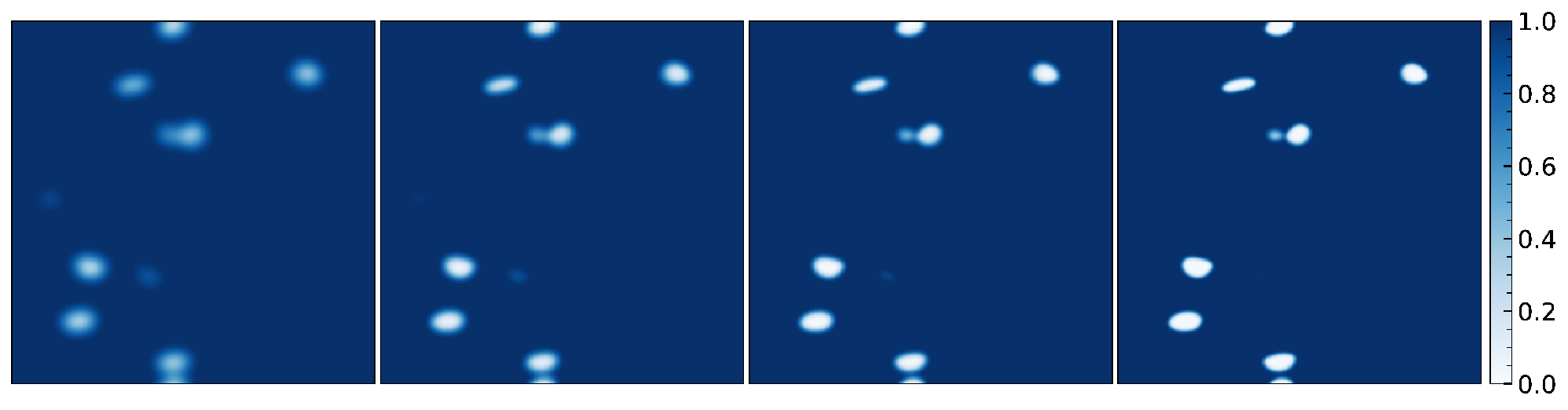}
    \includegraphics[width=1\linewidth]{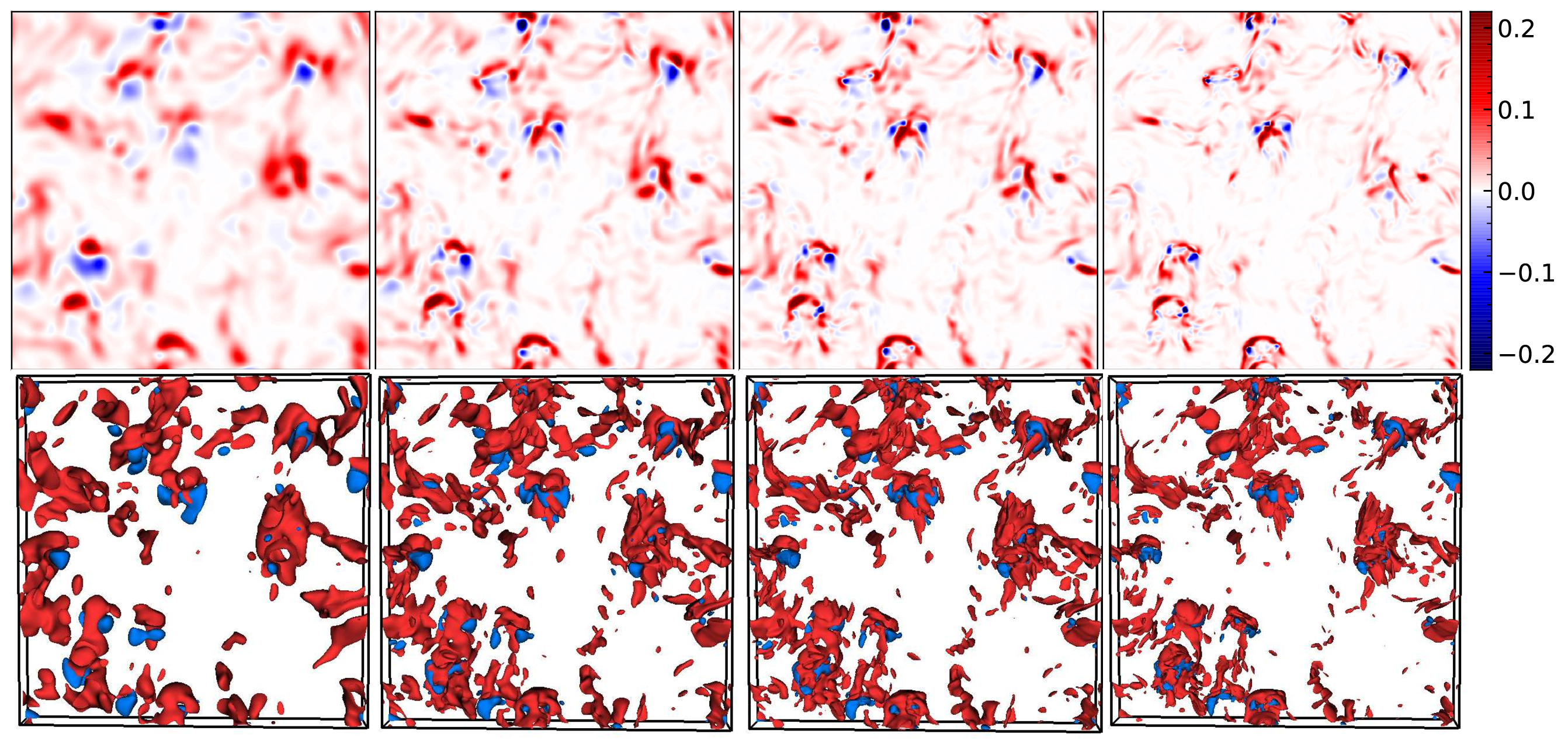}
    \caption{\label{flx_snap_R3} Buoyancy driven flow in presence of
      stirring ($b=0.13,{\tt R3}$).  The pseudocolor plot of the filtered
      indicator function $c$ (top panel) and the local nonlinear 
      flux $\piK/\max(\pi_{\kd})$ (middle panel) in the $y=L/2$
      plane. Constant-$\piK$ isosurfaces for $|\piK|=0.03 \max(\pi_{\kd})$  in a slab
      $L\times 2d\times L$ around the $y=L/2$ plane (bottom
      panel). The filter wavenumber (scale) is increased (decreased) from
      left to right $K/\kd=0.6, 1.0, 1.4$ and $2.2$.}
\end{figure}

\begin{figure}
    \centering
    \includegraphics[width=0.6\linewidth]{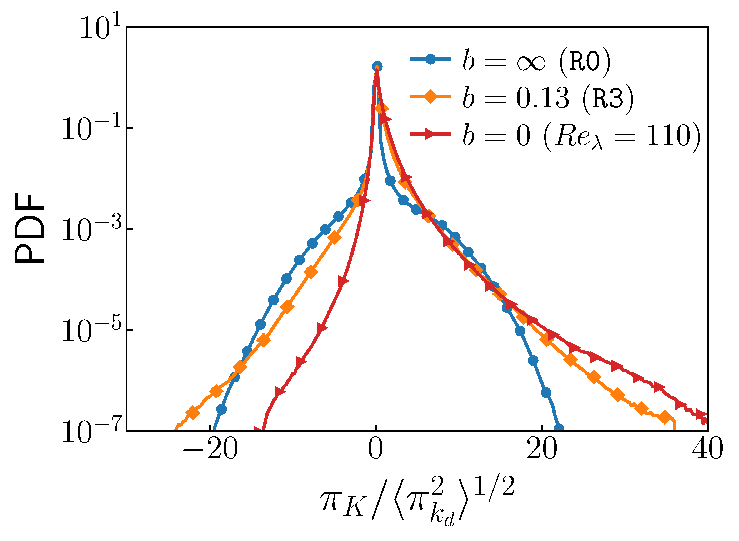}
    \caption{\label{pdf_flux_R3} The PDF of the scaled nonlinear flux $\piK/\braket{\piK^2}^{1/2}$
for different values of $b$, and with $K=\kd$.}
\end{figure}

The situation is more complex in presence of stirring, as now both the large scale 
forcing as well as the wake of the bubble creates complex spatio-temporal pattern 
for $\piK(\xx)$ with regions of downscale and upscale transfer (see \Fig{flx_snap_R3}).  
In \Fig{pdf_flux_R3} we plot the PDF of $\piK(\xx)$ with $K=\kd$ for $b=0,0.13$, and $b=\infty$. 
For all the cases we observe that the PDF is positively skewed confirming a net positive flux of energy. 
The skewness of the PDF for $b=0$ is nearly $1.3$ times larger than the $b=\infty$, 
indicating presence of stronger inverse energy transfers in buoyancy driven bubbly flows in 
comparison to homogeneous, isotropic turbulence. 
This is further verified by noting that the skewness for $b=0.13$, where both stirring and 
buoyancy driven bubbles generate turbulence, is smaller than the case with $b=0$.

\subsection{Total energy budget}
Using \eqref{ns1} we obtain the steady-state the total energy budget equation as 
\begin{align}
\epg+\ep=\epn
\label{eq:ebud}
\end{align}
 i.e., energy injected by buoyancy and stirring is dissipated by
 viscosity. Using table~\ref{tab:runs}, \eqref{eq:ebud} is easily verified.

In this section, we study the contribution to the total budget from each of the
phases. The two phases are characterized by the indicator function $c$ which
takes value $1$ in the liquid phase, $0$ inside the bubble and an intermediate
value at the interface. In a DNS of two-phase flows, usually, the interface is
diffused over to $3-4$ grid points. Thus, using $c$ to distinguish the phases
implies that the interface region contributes to both the phases.  In order to
avoid this conundrum, we construct a new indicator function $\cp$ such that the
interface points are included inside the bubble. \REM{The new function $\cp$ is
constructed by expanding $c$ radially by $0.16d$, where $d$ is the bubble diameter.}
{To construct $c^\prime$ we first initialize it to be the same as $c$.
  The points which lie closest to $c^\prime=1/2$ contour are identified as
  bubble interface points. For points where $c^\prime<1/2$, $c^\prime$ is set to zero and it is unity
  outside. Next we set $c^\prime=0$ at all points that are within a distance of $0.16d$
  from the interface points. This completes the procedure of generating an inflated region around each
  bubble.}

Henceforth we shall use the term bubble to indicate the regions where $\cp=0$.
Using $\cp$ we define the net injection and dissipation rates in the liquid as:
\begin{subequations}
    \begin{align}
	    \epn_{\rm l} &= 2 \nu\bra{\cp \mS : \mS  }\/, \\
	    \epg_{\rm l} &= \bra{ \cp\uu\cdot \FF^{\rm g} }\/,\\
	    \ep_{\rm l} &=\bra{ \cp\uu \cdot \FF^{\rm s} }\/.
    \end{align}
    \label{eq:liq_bud}
\end{subequations}
The contribution from the bubble phase can be obtained by subtracting
the contribution from the liquid phase from the total, for instance, 
dissipation rate in the bubble phase is $\epn_{\rm b} = \epn - \epn_{\rm l}$.

\begin{figure}
  \centering
      \includegraphics[width=0.48\linewidth]{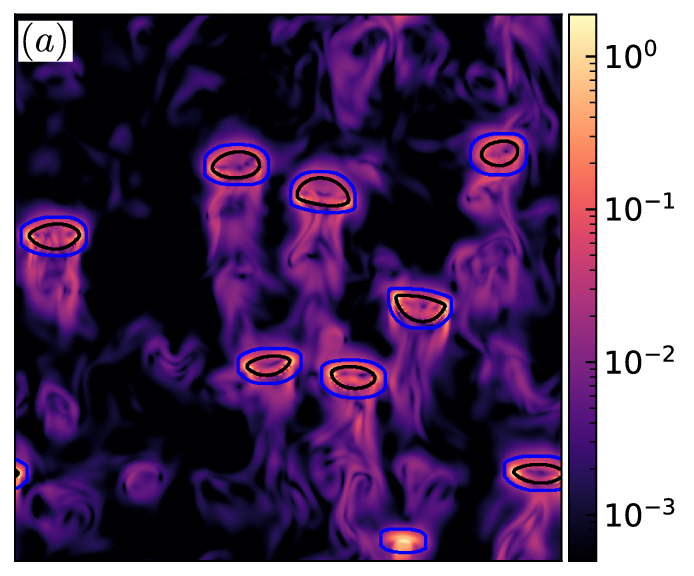}
      \includegraphics[width=0.48\linewidth]{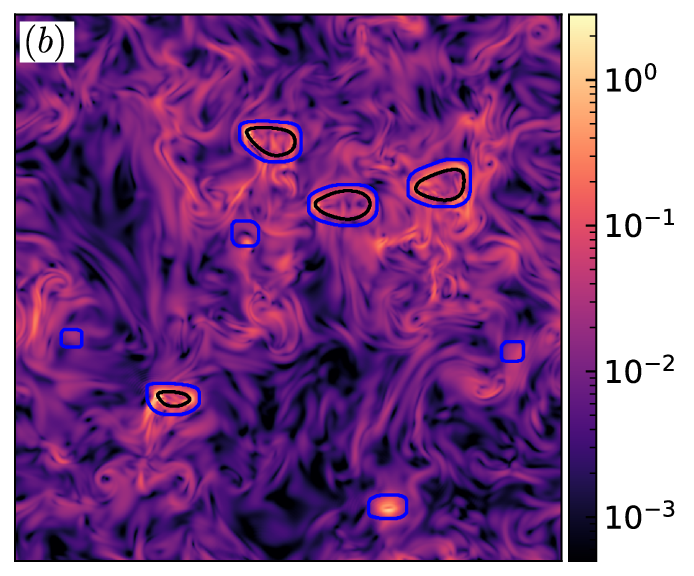}
      \caption{\label{fig:visc_snap} Pseudocolor plot of the local
        dissipation $\epn_{\rm loc}$ in $y=L/2$ plane for (a) $b=\infty$ (run
        ${\tt R0}$), and (b) $b=0.13$ (run {\tt R3}). The black line
        represents the bubble interface ($c=0$ contour), and blue line
        indicates the contour $c'=0$. }
\end{figure}

In \subfig{fig:visc_snap}{a,b} we show the pseudocolor plot of the local viscous dissipation 
$\epn_{\rm loc}(\xx)=2 \nu \mS:\mS$.
For the case with no stirring, $b=\infty$, the dissipation is strongly 
concentrated inside and in the wake of the bubbles, 
whereas when stirring is present, $b=0.13$, strong
dissipation is also observed in the liquid phase away from the bubbles. 

\begin{figure}
    \centering
    \includegraphics[width=0.45\linewidth]{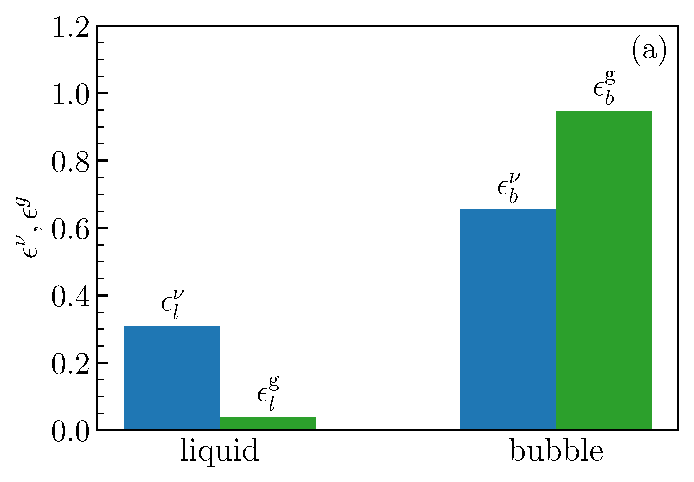}
    \includegraphics[width=0.45\linewidth]{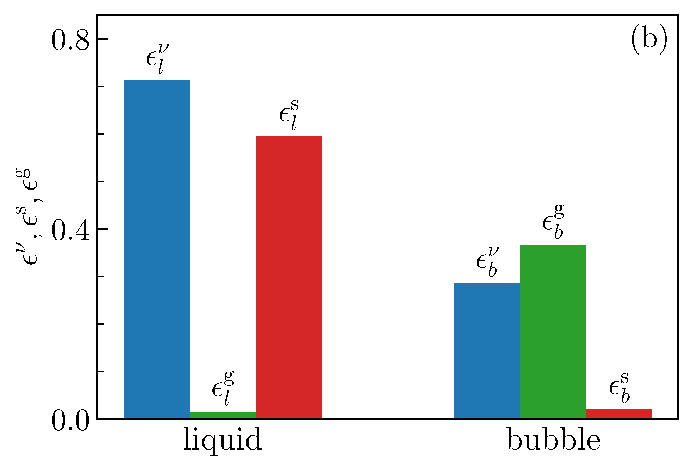}
     \caption{The dissipation and injection rates in the steady state
       evaluated in the liquid, and the bubble phase for (a) $b=\infty$ (run
       ${\tt R0}$) and (b) $b=0.13$ (run ${\tt R3}$). The ordinate in both
       the figures is normalized by $\epn$}
    \label{fig:ph_fp00}
\end{figure}

In \subfig{fig:ph_fp00}{a} we look at the balance between energy injection
and dissipation in each phase for the case of no stirring, $b=\infty$. 
In the liquid phase, viscous dissipation $\epn_{\rm l}$ far exceeds energy injected due to 
buoyancy $\epg_{\rm l}$, whereas in the bubble phase the situation is reversed.
Note that  the overall viscous dissipation inside the bubble phase is larger than the
overall dissipation in the liquid phase.  

We can now summarise the flow of energy completely for the case of no
stirring, $b=\infty$. Buoyancy force injects energy
at the scale of the bubbles, largely in the gas phase. A large
fraction of this energy is dissipated within the bubble itself. Rest
of it is transferred to the liquid phase by bubble-liquid
interaction. Both the nonlinear flux and the
flux due to the surface tension cascades this energy to smaller and
smaller scales in the fluid.  Energy dissipation happens in both the
gas and liquid phase starting from the scale of bubble down to the
smallest scales.

We next plot the injection and dissipation rates obtained for
different phases for the case $b=0.13$ in \subfig{fig:ph_fp00}{b}.
Here, we find that dominant energy injection is due to the stirring.
This appears largely in the liquid phase.  The net energy dissipated
in the liquid phase exceeds the energy injected  by
stirring due to the additional energy transfer from the bubble phase
to the liquid phase. In the bubble phase energy is injected by the
buoyancy forces.  Most of this energy is dissipated in the bubble
phase, but as pointed out above, a part of it is also transferred to
the liquid phase.

\section{Conclusion}
We conduct a DNS study of buoyancy-driven bubbly flow in the presence
of large-scale stirring. We investigate the statistical properties of
the flow and compare our findings with the experiments. Our key
results are summarised below:
\begin{enumerate}
\item The rise velocity of a bubble in the suspension reduces, and the
liquid velocity fluctuations are rendered isotropic on increasing the
stirring intensity.
\item Consistent with experiments \citep{lance_1991,vivek_2016}, we
find the energy spectrum shows a Kolmogorov scaling for $k \ll \kd$
and a pseudo-turbulence scaling -- $E(k) \sim k^{-3}$ -- for $k \gg
\kd$.
\item We rationalize the scaling observed
in the energy spectrum by using a scale-by-scale energy budget analysis. 
For $k\ll \kd$, energy flux is the dominant energy transfer mechanism
although viscous dissipation is effective for all scales $k < \kd$. 
The balance of net production with viscous dissipation leads to the 
pseudo-turbulence scaling for $k\gg \kd$. 
\end{enumerate}

We want to emphasize that although we study turbulence modulation by
weakly buoyant bubbles, the statistical properties of the flow are in
qualitative agreement with the experiments
\citep{lance_1991,vivek_2016,sali_2020}.  Therefore, we believe that
the energy transfer mechanisms discussed in our study should also
apply to the experimental scenario of high density and viscosity
contrast; our previous study \citep{pandey_2020} already verified this
in the absence of stirring.

However, we expect that the details of the wake structure in the
vicinity of the bubble would depend on the density and viscosity
contrast. How relevant is this for the energy transfer mechanism that
we have proposed remains to be investigated. We hope that our results
will motivate further investigations in this direction.

\appendix
\section{Boussinesq approximated Navier-Stokes equations} \label{app:bouss}
In this section we derive the Boussinesq approximate equations \eqref{ns1} starting from the 
 following  multiphase Navier-Stokes equations  \citep{pandey_2020}:
\begin{align}
\rho(c) D_t \uu = \nabla \cdot (\mu(c) \mS) + {\bm f}^{\sigma} + {\bm f},
\label{eqa:gen}
\end{align}
where the density field 
\begin{align}
\rho(c)=\rhof c  + \rhob (1-c),
\label{eqa:rho}
\end{align}
the dynamic viscosity field $\mu(c)=\muf c + \mub (1-c)$, $\rhof (\muf)$ is the
density (viscosity) of fluid phase, $\rhob (\mub)$ is the density (viscosity) of the
bubble phase, ${\bm f}^{\sigma}$ is the surface tension force,  the external force ${\bm f} \equiv [\rho(c) {\bm a}- \overline{\rho(c) {\bm a}}]$, ${\bm a}$ is the acceleration, and in this section $\overline{(\cdot)}$ denotes spatial averaging.   Note that as we work with periodic boundaries, our choice of external force ensures that no net momentum is added to the flow.

We assume small density contrast ($\At\ll 1$) and identical dynamical viscosity  of the  two phases ($\muf/\mub=1$). Thus we invoke Boussinesq approximation,  whereby $\rho(c)$ on the left-hand side of \eqref{eqa:gen} is replaced by the average density 
\begin{align}
\rhoa\equiv \overline{\rho(c)} = (\rhof-\rhob) c_a + \rhob \approx (\rhof + \rhob)/2.
\label{eqa:rhoa}
\end{align}

The above assumptions drastically simplify \eqref{eqa:gen} to give,
\begin{align}
D_t \uu = \nu \nabla^2\uu  + \Fs + {\bm F},
\end{align}
where $\Fs={\bm f}^{\sigma}/\rhoa$ and ${\bm F}={\bm f}/\rhoa$. The above equation is identical to the Boussinesq equation \eqref{ns1} that we use. Next we derive the buoyancy and the turbulent stirring force in the Boussinesq regime.

Using the definitions \eqref{eqa:rho} and \eqref{eqa:rhoa}  in ${\bm F}$ we get,
\begin{align}
{\bm F} = \left[ 1 - {(\rhof-\rhob)\ca\over \rhoa} \right] ({\bm a} - \overline{\bm a}) + {(\rhof-\rhob)\over \rhoa} (c {\bm a} - \overline{c {\bm a} }).
\label{eq:force}
\end{align}

When ${\bm a}={\bm g}$, the first term on the right hand side of \eqref{eq:force} is zero and we obtain the buoyancy force  
\begin{align}
{\bm F}^g = {(\rhof-\rhob)\over \rhoa} (c - \ca) {\bm g} \approx 2 \At (c-\ca) {\bm g}.
\end{align}

On the other hand for turbulence stirring, we use an acceleration field with $\overline{{\bm a}} =0$. Therefore, \eqref{eq:force} simplifies to:
 \begin{align}
\Ftrb = \left[ 1 - {(\rhof-\rhob)\ca\over \rhoa} \right] {\bm a} +
{(\rhof-\rhob)\over \rhoa} (c {\bm a} - \overline{c {\bm a} }).
\end{align}
In the Boussinesq regime, $(\rhof-\rhob)/\rhoa \ll 1$ and we get $\Ftrb={\bm a}$ to the leading order, i.e., the stirring force is applied irrespective of the phase or the indicator function.  In the main manuscript we choose $\rhoa=1$ everywhere.

\section{Resolution test} \label{app:sec2}
\begin{figure}
    \includegraphics[width=0.48\linewidth]{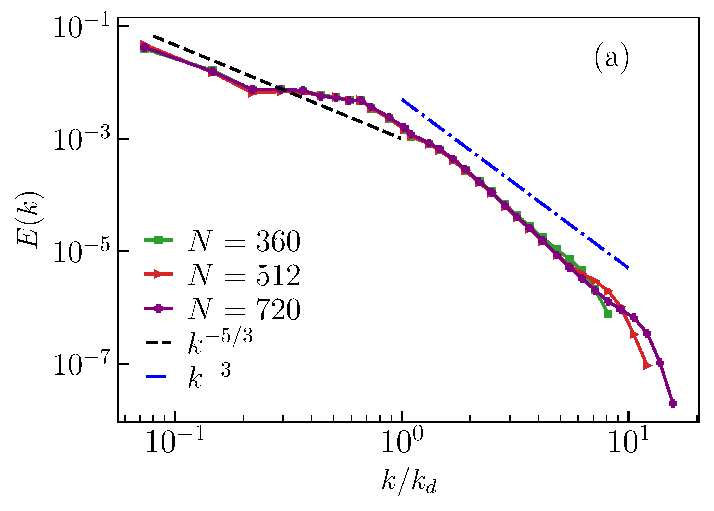}
    \includegraphics[width=0.48\linewidth]{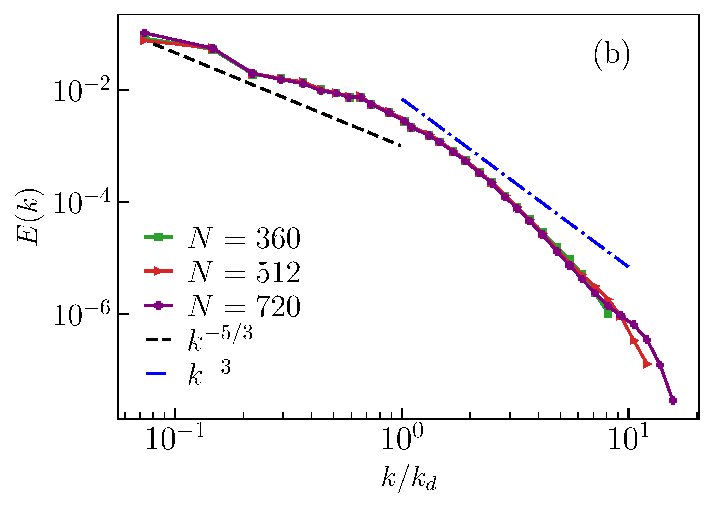}
    \caption{\label{restest} The kinetic energy spectra for $\tt{R1}$
      (left) and $\tt{R3}$ (right) at resolutions $N=360,512$ and $720$.}
  \end{figure}

    To study grid convergence, we conduct DNS of
    turbulent bubbly flows for our runs ${\tt R1}$ and ${\tt R3}$ with
    increasing grid-resolution $N=360, 512$, and $720$.
    The plot of the energy spectrum \subfig{restest}{} clearly shows that
    even with $N=360$, the inertial range as well as the $k^{-3}$ scaling of
    pseudo-turbulence are well-captured. However, as expected, on increasing
    the grid-resolution the range of $k^{-3}$ scaling  obtained due the balance
    of net production with viscous dissipation  extends.
    The departure from the $k^{-3}$ scaling around $k\approx k_{\rm max}$ is
   an artifact of finite resolution.

\section{Energy budget using sharp filter} \label{app}
We now present the result of the scale-by-scale energy budget analysis
obtained by using a sharp low-pass filter instead of the Gaussian
filter. The low-pass filtered velocity field for a sharp filter is
defined as \citep{frisch,verma_2019,pandey_2020}:
\begin{equation}
 \uK(\xx) =  \sum_{q \le K} \bm{u}_{\bm q}\exp(i\bm{q}\cdot\bm{x}).
\end{equation}

\begin{figure}
  \begin{center}
      \includegraphics[width=0.45\linewidth]{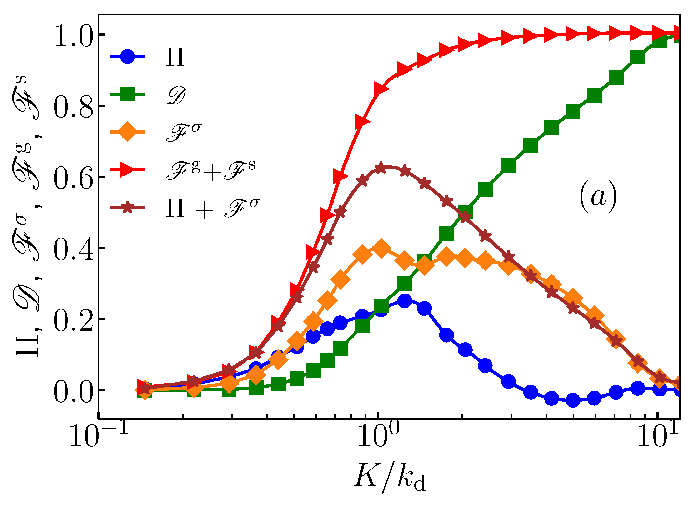}
      \includegraphics[width=0.45\linewidth]{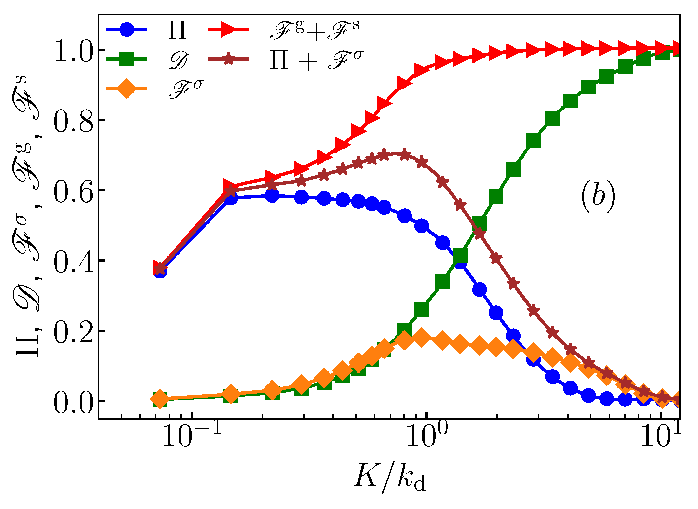}
      \caption{\label{fig:flux_direct} Scale-by-scale energy budget: plot of the energy flux $\PiK$, cumulative 
      viscous dissipation $\DK$, the surface tension contribution $\FsK$, the 
      cumulative energy injected due to buoyancy $\FgK$, and the energy injected due
  to turbulent forcing $\FturbK$ for  (a) $b=\infty$, and  (b)
  $b=0.13$. In both the panels we normalize the ordinate by the 
        viscous dissipation $\epn$.}
  \end{center}
\end{figure}

In \subfig{fig:flux_direct}{a} we show the scale-by-scale budget
obtained for the case $b=\infty$ and in \subfig{fig:flux_direct}{b} we
plot the budget for $b=0.13$. By comparing with \Fig{fig:flux}, it is
clear that the choice of filtering does not qualitatively change the
scale-by-scale energy budget. Our observations are consistent with the
recent finding of \citet{chibbaro_2020} who did a similar comparison
for homogeneous, isotropic turbulence.

\section*{Author contributions}
V.P. performed  the simulations. All
authors contributed equally to analysing data and reaching conclusions, and in writing the paper.

\section*{Funding}
This work was supported by the Department of Atomic Energy (DAE), India under
Project Identification No. RTI 4007,  DST (India) Project Nos. ECR/2018/001135 and
DST/NSM/R\&D\_HPC\_Applications/2021/29, and the Swedish Research Council Grant No.
638-2013-9243 as well as 2016-05225.

\section*{Declaration of interests}
The authors report no conflict of interest.


\input{main_document.bbl}
\end{document}

%% file: symdef.inc
\def \cp    {c^{\prime}}

\def \muf   {\mu_{\rm f}}
\def \mub   {\mu_{\rm b}}
\def \piK   {\pi_{ K}}
\def \keta  {k_{\eta}}
\def \ua    {u^{\alpha}}
\def \ub    {u^{\beta}}
\def \dela  {\partial_{\alpha}}
\def \delb  {\partial_{\beta}}
\def \ab    {\alpha\beta}
\def \mT    {\mathsfbi{T}}
\def \mS    {\mathsfbi{S}}
\def \mSK    {\mathsfbi{S}_{ K}}
\def \PiK   {\Pi_{ K}}
\def \FsK   {\mathscr{F}^\sigma_{ K}}
\def \Fs    {\bm{F}^\sigma}
\def \FgK   {\mathscr{F}^{\rm g}_{ K}}
\def \Fg    {\bm{F}^{\rm g}}
\def \DK    {\mathscr{D}_{ K}}
\def \FturbK {\mathscr{F}^{\rm s}_{ K}}
\def \Ftrb {\bm{F}^{\rm s}}
\def \uK    {\bm{u}^{<}_{ K}}
\def \GK    {G_{K}}
\def \qq    {\bm{q}}
\def \kd    {k_{\rm d}}
\def \Vzero {V_0}
\def \uzero {u_0}

\def \kk   {\bm{k}}
\def \FFh {\hat{\bm{F}}}
\def \uuh {\hat{\bm{u}}}
\def \grad {\bm{\nabla}}

\def \VV    {\bm{V}}
\def \tauL  {\tau_{\rm L}}
\def \uu   {\bm{u}}
\def \Nb   {N_{\rm b}}
\def \ep   {\varepsilon^{\rm s}}
\def \epg  {\varepsilon^{\rm g}}
\def \epn  {\varepsilon^{\nu}}
\def \kk   {\bm{k}}
\def \nhat {\bm{\hat{n}}}
\def \xx  {\bm{x}}
\def \FF  {\bm{F}}

\def \rhoa {\rho_{\rm a}}
\def \ca {c_{\rm a}}
\def \muf {\mu_{\rm f}}
\def \mub {\mu_{\rm b}}

\def \at {a_{t}}

\def \rhof {\rho_{\rm f}}
\def \rhob {\rho_{\rm b}}

\def \rhoa {\rho_{\rm a}}

\def \uu  {{\bm u}}

\def  \xx  {{\bm x}}
\def  \XX  {{\bm X}}

\def  \VV  {{\bm V}}

\def \kinj {k_{\rm inj}}

\def \grad {{\bm \nabla}}

\newcommand{\braket}[1]{\langle #1\rangle}
\newcommand{\paren}[1]{\left(#1\right)}
\newcommand{\bra}[1]{\left\langle #1\right\rangle}
\newcommand{\fil}[1]{\left(#1\right)^{<}_{K}}
\def \Rey  {\mbox{Re}_{\lambda}}
\def \Ga  {\mbox{Ga}}

\def \At  {\mbox{At}}
\def \Bo  {\mbox{Bo}}

\def\i{{\rm i}}

\def\d{{\rm d}}

\newcommand{\beq}{\begin{equation}}
\newcommand{\eeq}{\end{equation}}

\newcommand{\symb}[1]{\text{\sout{\ensuremath{~#1}~}}}
\newcommand{\REM}[1]{{}}
\newcommand{\eq}[1]{~(\ref{#1})}
\newcommand{\Eq}[1]{Eq.~(\ref{#1})}
\newcommand{\Fig}[1]{figure~(\ref{#1})}
\newcommand{\subfig}[2]{figure~(\ref{#1}#2)}